\documentclass[useAMS,usenatbib,usegraphicx]{mn2e}

\usepackage{times}
\usepackage{BibDef}
\usepackage{hyperref}
\usepackage{amssymb}
\usepackage{float}
\usepackage{breakurl}

\newcommand{\de}{{\rm d}}
\newcommand{\deriv}[2]{ \frac{{\rm d} #1}{{\rm d} #2} }
\newcommand{\pderiv}[2]{ \frac{\partial #1}{\partial #2} }

\usepackage{color}

\title[Radiation-dominated winds and quasi-stars]{Bright vigorous winds as signposts of supermassive black hole birth}

\author[Fiacconi \& Rossi]{Davide Fiacconi$^{1,}$\thanks{E-mail: fiacconi@physik.uzh.ch} \& Elena M. Rossi$^{2,}$\thanks{E-mail: emr@strw.leidenuniv.nl}\\
$^{1}$Center for Theoretical Astrophysics and Cosmology, Institute for Computational Science, University of Zurich, Winterthurerstrasse 190,\\
CH-8057 Z\"{u}rich, Switzerland\\
$^{2}$Leiden Observatory, Leiden University, PO Box 9513, 2300 RA, Leiden, the Netherlands
}

\begin{document}

\date{\today}

\pagerange{\pageref{firstpage}--\pageref{lastpage}} \pubyear{2015}

\maketitle

\label{firstpage}


\begin{abstract}
The formation of supermassive black holes is still an outstanding question. In the {\it quasi-star} scenario, black hole seeds experience an initial super-Eddington growth, that in less than a million years may leave a $10^4-10^5$~M$_{\sun}$ black hole at the centre of a protogalaxy at $z \sim 20-10$. 
Super-Eddington accretion, however, may be accompanied by vigorous mass loss that can limit the amount of mass that reaches the black hole.
In this paper, we critically assess the  impact of radiative driven winds, launched from the surface of the massive envelopes from which the black hole accretes. 
Solving the full wind equations coupled with the hydrostatic structure of the envelope, we find mass outflows with rates between a few tens and $10^4$~M$_{\sun}$~yr$^{-1}$, mainly powered by advection luminosity within the outflow. 
We therefore confirm the claim by \citet{dotan+11} that mass losses can severely affect the black hole seed early growth within a quasi-star.
In particular, seeds with mass  $>10^4$~M$_{\sun}$ can only form within mass reservoirs $ \gtrsim 10^7$~M$_{\sun}$, unless they are refilled at huge rates ($ \gtrsim 100$~M$_{\sun}$~yr$^{-1}$). 
This may imply that only very massive halos ($>10^9$~M$_{\sun}$) at those redshifts can harbour massive seeds. 
Contrary to previous claims, these winds are expected to be relatively bright ($10^{44}-10^{47}$~erg~s$^{-1}$), blue ($T_{\rm eff} \sim 8000$ K) objects, that while eluding the Hubble Space Telescope, could be observed by the James Webb Space Telescope. 
\end{abstract}



\section{Introduction}

Supermassive black hole formation is an outstanding  question in astrophysics.
The compelling evidence that links supermassive black holes' evolution to that of their
host galaxies (e.g. \citealt{magorrian+98, ferrarese+00, tremaine+02,
marconi+03,haring+04,gultekin+09,mcconnell+13}), strongly suggests that the
answer must be sought in the broader context of galaxy assembly.

A few different formation scenarios have been proposed. Supermassive black hole seeds might have a
classical stellar origin as the left over of the first generation of stars
(e.g. \citealt{madau+01,volonteri+03,tanaka+09}) or they might represent the outcome of the core collapse
of primordial nuclear star clusters (e.g. \citealt{quinlan+90,devecchi+09,davies+11,devecchi+12,lupi+14}).  
Both proposals, however, face difficulties in explaining the few observational constraints available,
namely the early occurrence of massive ($\gtrsim 10^{9}-10^{10}$~M$_{\sun}$) accreting black holes in
$z \sim 6-7$ quasars (\citealt{fan+06,mortlock+11}; but see also \citealt{treister+13}).  The main
reason is that both scenarios tend to predict small seeds (up to $\sim 1000$~M$_{\sun}$ at most)
which are unlikely to grow fast enough to power those high-$z$ quasars, unless sustained
super-Eddington accretion is advocated
(e.g. \citealt{johnson+07,pelupessy+07,milosavljevic+09,alexander+14,madau+14,volonteri+15}).
 
Although these observational constraints do not necessarily rule out those mechanisms on a physical
base, they may more easily be explained by the so called ``direct collapse'' scenario
(e.g. \citealt{bromm+03,begelman+06,lodato+06,dijkstra+08,begelman+09,latif+13,mayer+15}).  This
latter envisages a large mass of pristine gas ($\sim 10^6-10^7$~M$_{\sun}$), promptly accumulated at
the centre of a galaxy-size halo on (sub)parsec scales.  A large fraction of it would rapidly
($<10^6$ yr) form a massive seed ($10^{4}-10^{6}$~M$_{\sun}$), directly at the centre of a galaxy at $z\sim 15$.
Attractive as it is, this process is far from being proved and at least two major steps require further investigations.

  Although in principle there is plenty of gas available at high redshift and cold flows have been
  shown to be effective in bringing that gas down to the centre of (massive) halos
  \citep{dimatteo+12}, the conditions to avoid substantial fragmentations and to overcome the
  centrifugal barrier are not fully understood yet.  Several possibilities have been
  discussed, such as the dissociation of H$_2$ molecules by Lyman-Werner ionising radiation coming
  from nearby, star-forming galaxies in order to avoid cooling and fragmentation
  \citep{ferrara+13,dijkstra+14}.  Another possibility is the onset of supersonic turbulence and the
  removal of angular momentum due to non-axisymmetric perturbations and gravitational torques during
  the collapse of the halo \citep{begelman+09,choi+13,choi+15}, or at the centre of major merger remnants between
  rare and massive galaxies at high redshift \citep{mayer+10, mayer+15}.  

 The second issue is  how to actually form a black hole and what is its initial mass. 
The answer may vary according to the physical properties of the assembled mass.
When more than $\sim 10^8$~M$_{\sun}$  can be rapidly piled up, 
the resulting structure likely becomes dynamically unstable (even if rotating) and relativistic radial
 instability can lead to implosion and direct black hole formation
  \citep{fowler+66,baumgarte+99,shibata+02}.  However, forming such a structure requires rather
  extreme conditions (e.g. a major merger; \citealt{mayer+15}).  With relatively milder accretion
rates, nuclear burning can start at the centre of a convectively stable object, i.e a supermassive
star \citep{begelman+10}.  After a million years, the core that could not convectively acquire fresh
hydrogen collapses to form a stellar size ($\sim 100$~M$_{\sun}$) black hole.  Highly optically thick
gas keeps however falling onto the newly born black hole, with enough angular momentum to be able to
generate accretion power.  This energy feedback inflates the innermost part of this inflow, creating a
{\it quasi-star}: a massive, slowly rotating envelope, sustained against its own gravity by the black
hole accretion power \citep{begelman+08,begelman+10, volonteri+10,ball+11,dotan+11}.  At quasi-star
centres, the embryo black holes may accrete at a super-Eddington rate, as energy is transported
outward by convection (not by radiative diffusion) through the envelope.  The initial expectation was
that seeds of $\sim 10^4-10^5$~M$_{\sun}$ may easily grow in $\lesssim 1$~Myr \citep{begelman+08}.
After this time, the envelope would be definitively dispersed and accretion would proceed at an
Eddington limited fashion directly from the protogalactic disc.

This result was questioned by \citet{dotan+11}, that tried to quantify the impact of radiative driven
winds from the surface of these weakly bound envelopes.  They found that in a large part of the black
hole mass-envelope mass parameter space, winds can be so powerful that the envelope evaporates before
the black hole is able to double its mass.  This effect would greatly limit the
number of protogalaxies in whose centre the conditions are prone to massive $>10^4$~M$_{\sun}$ seed
formation.  Moreover, because in their model most of the radiation energy in diffusive luminosity is
converted into kinetic energy of the wind, quasi-stars would be very dim objects, virtually
undetectable. In that paper, however, the radiative driven wind was not modelled
solving the full equations of motion and in particular, the advection energy term was neglected. 

We therefore set out to critically reconsider continuum driven winds from the surface of radiation-dominated
objects. In fact, we find that the advection term has an important dynamical role as the main driver of the wind.
We then explicitly consider quasi-stars and calculate mass loss rates and photospheric properties
and we assess their detectability with Hubble and the James Webb Space Telescopes (JWST).  Our result is that
although winds are still a major limit for black hole growth, photospheric luminosities in the wind
ensure possible bright targets ($10^{44}-10^{46}$~erg~s$^{-1}$) for JWST.

This paper is organised as follows: in Section \ref{sec_2} we describe our model of radiation-dominated 
wind and we discuss the main properties; in Section \ref{sec_3} we couple this wind
prescription to the hydrostatic envelope predicted for quasi-stars, finding equilibrium solution and
discussing their evolution.  Section \ref{sec_4} is devoted to simple predictions regarding the
observability of quasi-stars by current and future space-based telescopes.  We discuss and
summarise our main findings in Section \ref{sec_5}, addressing the main limitations
 of our work and future steps.


\section{The wind model} \label{sec_2}


\subsection{Equations and general properties} \label{subsec_equation}

We consider a stationary, spherically-symmetric, radiation-dominated wind launched by a
non-rotating, stellar-like object of mass $M_{\star}$ from a spherical surface of radius $R_{\star}$,
that represents the base of the wind\footnote{In the following, we will always use the subscript
  $_\star$ to indicate quantities evaluated at $R_{\star}$}.
By radiation-dominated, we mean that the contribution of the gas pressure $p_{\rm gas}$ is assumed to be
negligible compared to the radiation pressure $p_{\rm rad}$, i.e. $p_{\rm gas} / p_{\rm rad} \ll 1$.
This assumption allows us to neglect the presence of $p_{\rm gas}$ in the following calculations (and we simply write $p \equiv p_{\rm rad}$), but it requires at the same time that the wind is launched from a radiation-dominated object (as we consistently show for quasi-stars in Section \ref{sec_3}).
Just outside $R_{\star}$, the gas
is assumed to be initially optically thick and interacts with radiation through a constant opacity
$\kappa$.
In this Section we implicitly assume the Thompson scattering opacity (though the specific value is in fact irrelevant for our results since we develop all our calculations in a dimensionless form), while in Section \ref{sec_3} we will adopt a temperature-dependent opacity law (see Section \ref{subsec_equation_qs}).
We stress, however, that the (dynamical) results presented in this Section are largely independent of the assumed opacity law, as also discussed in Section \ref{subsec:results_qs}.
We are interested in primordial composition objects, where line-driven interaction is negligible.
The equations that describe this system are similar to those used by several previous works about stellar
winds (and spherical accretion) in both the optically thin and optically thick regime
(e.g. \citealt{zytkow+72, begelman+78, begelman+79, kato+83, quinn+85}):
\begin{equation} \label{eq_mass_cons}
\dot{M} = 4 \pi r^2 \rho~v,
\end{equation}
\begin{equation} \label{eq_mom_cons}
\frac{1}{2} \deriv{v^2}{r} =  -\frac{G M_{\star}}{r^2} + \frac{\kappa L}{4 \pi r^2 c} ,
\end{equation}
\begin{equation} \label{eq_en_cons}
\dot{M} \frac{\de}{\de r} \left( \frac{v^2}{2} - \frac{G M_{\star}}{r} + \frac{p + U}{\rho} \right)  =  -\deriv{L}{r}.
\end{equation}
These equations determine the structure of the gas density $\rho$, the radial gas velocity $v$, the
luminosity carried by photons $L$, the (radiation) pressure $p$ and the (radiation) internal energy
density $U$ as a function of the spherical radius $r$ within the gravitation potential
$\Phi = -G M_{\star}/r$ induced by $M_{\star}$ outside $R_{\star}$.  The steady-state wind is
characterised by the constant outflow rate $\dot{M}$.  Equations \ref{eq_mass_cons}, \ref{eq_mom_cons}
and \ref{eq_en_cons} describe the conservation of mass, momentum, and energy, respectively.

Such a system of equations is not closed and several approaches can be used to close it to different 
degrees of approximation. \citet{shaviv+01b} and \citet{owocki+04} start from similar equations, except 
that they initially include the contribution of the gas thermal pressure to the momentum and energy 
conservation. Then, they simplify the system focusing on the supersonic branch, thus subsequently neglecting the gas 
pressure terms in the momentum equation (which brings it back to our same equation \ref{eq_mom_cons}) and 
the advection term $(p+U)/\rho$ (with the corresponding one due to gas pressure) in the energy equation.
This approximation leads to the great advantage that fully analytic solutions can be derived.
However, the limitation is that the lack of the advective term makes the behaviour of the wind insensitive to the local optical thickness.

Instead, we follow an approach similar to that used by e.g. \citet{quinn+85} and we explicitly include 
additional prescriptions to properly describe the behaviour of the wind in the extrema of very optically 
thin and optically thick regime, i.e. when the optical depth:
\begin{equation}
\tau(r) = \int_{r}^{+\infty} \kappa~\rho(x)~\de x,
\end{equation}
is either $\tau \ll 1$ or $\tau \gg 1$, respectively.  When the wind is optically thick, radiation and
matter can reach local thermodynamical equilibrium at the same temperature $T$ (which relates to the
energy density $U = 3 p = a T^4$, where $a$ is the radiation constant) and the gradient of the
radiation energy density is:
\begin{equation}
\deriv{U}{r} \Big|_{\tau \gg 1} = - \frac{3 \kappa \rho L}{4 \pi r^2 c}.
\end{equation}
On the other hand, local thermodynamical equilibrium may not be reached in the optically thin limit
and a unique temperature may not be a physically-motivated quantity.  In this case, the photons
carrying $L$ travel with roughly radial orbits and interact very little with matter, keeping $L$
almost constant (see e.g. \citealt{zytkow+72}).  Then, the radiation energy density decreases mostly
because of geometrical dilution in a progressively larger volume:
\begin{equation}
\deriv{U}{r} \Big|_{\tau \ll 1} = - \frac{L}{2 \pi r^3 c}.
\end{equation}
We follow \citet{quinn+85} in defining the total gradient of $U$ as the sum of the two limiting cases:
\begin{equation} \label{eq_grad_u}
\deriv{U}{r} = - \frac{L}{2 \pi r^3 c}~ f(\tilde{\tau}),
\end{equation}
where we define the function:
\begin{equation} \label{eq_f_tau}
f(\tilde{\tau}) = \frac{3 \tilde{\tau}}{2} + 1, \quad \tilde{\tau} \equiv \kappa~\rho~r.
\end{equation}
The ``effective'' opacity $\tilde{\tau}$ leads the gradient of $U$ to the right optically thin and
optically thick limits when $\tilde{\tau} \ll 1$ and $\tilde{\tau} \gg 1$, respectively.  However,
$\tilde{\tau}$ is just an approximation of the actual opacity $\tau$; the two are related by a
constant factor when $\tau$ is a power law and such a factor is close to 1 when $\tau \propto r^{-1}$.
Although we do not know a priory the relationship between $\tau$ and $\tilde{\tau}$, we demonstrate in
the following that $\tau \propto r^{-1}$ roughly holds and therefore $\tilde{\tau} \simeq \tau$ (see
Section \ref{sec_3}).

\begin{figure}
\begin{center}
\includegraphics[width=8cm]{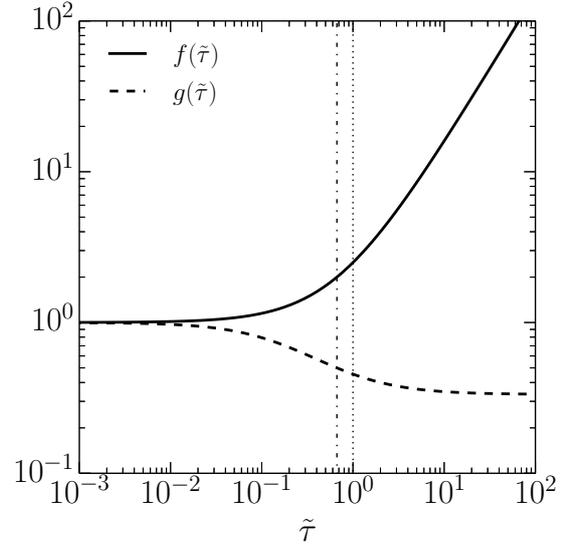}
\end{center}
\caption{The tick continuous and dashed lines show the behaviour of $f$ and $g$ as a function of 
$\tilde{\tau}$, respectively.
For reference, the vertical dotted and dashed lines mark the points $\tilde{\tau} = 1$ and 
$\tilde{\tau} = 2/3$, respectively.}
\label{fig_opt_depth_fac}
\end{figure}

Finally, we need to relate $p$ and $U$ to close the system of equations.  \citet{quinn+85} implicitly
assume that $p = U/3$ everywhere in the flow (see their equations 11b and 12).  This is correct in the
optically thick regime, but is not valid when the gas is optically thin.  Indeed, $U = p$ when the gas
is optically thin; this different relation between $p$ and $U$ is also responsible for the inexact
relation between the luminosity observed by an observer at infinity and by an observer comoving with
the flow, as reported by \citeauthor{quinn+85} (\citeyear{quinn+85}; see also Section
\ref{subsec_num_integr} and \citealt{cassinelli+73}).  In order to have a smooth transition between
the two regimes, similar to the case of the gradient of $U$ (see equation \ref{eq_grad_u}), we propose
the following functional form for the opacity-dependent ratio $p/U$:
\begin{equation} \label{eq_g_tau}
\frac{p}{U} \equiv g(\tilde{\tau}) \equiv \left( \frac{3 \tilde{\tau}}{2} + 1 \right) \left( \frac{9 \tilde{\tau}}{2} + 1 \right)^{-1}.
\end{equation}
Figure \ref{fig_opt_depth_fac} shows the behaviour of both $f(\tilde{\tau})$ and $g(\tilde{\tau})$.
The latter goes correctly from 1 (when the wind is optically thin) to 1/3 (when the wind is optically
thick).  However, the accuracy of both prescriptions is questionable around $\tilde{\tau} \sim 1$, because 
the actual form is largely arbitrary. 
We therefore compute and compare wind models, choosing different functional form for $f(\tilde{\tau})$
and $g(\tilde{\tau})$ and concluded that our results are not affected as long as the the limits are correct and the
transition occurs rapidly (over several $\tilde{\tau}$) around $\tilde{\tau}\sim 1$.

The equation of the conservation of energy can be directly integrated, becoming an algebraic equation for $L$:
\begin{equation} \label{eq_energy_int}
L(r) = \dot{E} - \dot{M} \left(\frac{v^2}{2} - \frac{G M_{\star}}{r} + (1+g(\tilde{\tau})) \frac{U}{\rho} \right),
\end{equation}
where we make use of $g(\tilde{\tau})$ explicitly and we introduce the total conserved luminosity
$\dot{E}$, which represents the constant of integration.  The system of equations that we finally solve (often dubbed as ``wind equations'' in the following) is composed of equations
\ref{eq_mass_cons}, \ref{eq_mom_cons}, \ref{eq_grad_u} and \ref{eq_energy_int}, coupled with the
definitions of $\tilde{\tau}$, $f(\tilde{\tau})$ and $g(\tilde{\tau})$.  The system has two
ordinary differential equations and two algebraic equations for the dependent variables $\rho$ (or
$\tilde{\tau}$), $v$, $L$ and $U$ as a function of $r$.

At this point, it is convenient to introduce new dimensionless variables.
We define the new velocity variable $w = v^2/v_{\rm esc}^2$, where
$v_{\rm esc}^{2} = 2 G M_{\star} / R_{\star}$ is the escape velocity from the base of the wind, the
new radiation energy density variable $u = U \kappa R_{\star}^2 / (G M_{\star})$, and the new
luminosity variable $\Gamma = L/L_{\rm Edd}$.  $\Gamma$ is the Eddington ratio and the Eddington
luminosity $L_{\rm Edd}$ is defined as:
\begin{equation} \label{eq_eddington}
L_{\rm Edd} = \frac{4 \pi c G M_{\star}}{\kappa} = 1.26 \times 10^{38}~\tilde{\kappa}^{-1}~m_{\star}~{\rm erg~s^{-1}},
\end{equation}
where $\tilde{\kappa}$ is the opacity in units of the electron scattering opacity $\kappa_{\rm es} = 0.35$~cm$^{2}$~g$^{-1}$ (assuming primordial abundances) and $m_{\star}$ is the stellar mass $M_{\star}$ in units of solar masses.
The independent variable $r$ can also be transformed into $x = 1 - R_{\star}/r$, such that the interval $r \in [R_{\star},+\infty)$ is mapped into $x \in [0,1)$.
We can first express equation \ref{eq_mom_cons} with the new variables as:
\begin{equation}\label{eq_w}
w' = \Gamma - 1,
\end{equation}
where here and in the following $' = \de / \de x$.
The gradient of the radiation energy density $u$ becomes:
\begin{equation} \label{eq_u}
u' = -2~(1-x)~\Gamma~f(\tilde{\tau}),
\end{equation}
where $f(\tilde{\tau})$ is defined in equation \ref{eq_f_tau} and:
\begin{equation} \label{eq_tau}
\tilde{\tau} = \frac{\alpha \beta}{w^{1/2}} (1 - x),
\end{equation}
is the definition of $\tilde{\tau}$ using our dimensionless variables.
We introduce the two factors $\alpha$ and $\beta$; $\alpha$ is a dimensionless expression for $\dot{M}$ in terms of the trapping radius $R_{\rm tr}$ \citep{begelman+78,begelman+79}:
\begin{equation}\label{eq:rtr}
\alpha = \frac{R_{\rm tr}}{R_{\star}} = \frac{\kappa \dot{M}}{4 \pi c R_{\star}}.
\end{equation}
The trapping radius is where the diffusion time scale for photons is equal to the dynamical time of the outflowing wind, $R_{\rm tr} (\tilde{\tau}/c) \approx R_{\rm tr}/v$, which implies that $\tilde{\tau} \approx c/v$ at $R_{\rm tr}$.
We will see in the following that within this radius, since the radiation is \emph{trapped}, the luminosity  transported by diffusion becomes subdominant with respect to the energy advected within the flow.
The parameter $\beta$ is a dimensionless factor depending on the properties of the star only:
\begin{equation}
\beta \equiv \frac{c}{v_{\rm esc}} \approx 486~m_{\star}^{-1/2}~r_{\star}^{1/2},
\end{equation}
where $r_{\star}$ is the stellar radius in units of solar radii.
$\beta$ appears naturally from the normalisation of the optical depth and measures the deepness of the gravitational potential well of the parent star.
It depends on the stellar properties $R_{\star}$ and $M_{\star}$ only and therefore, $M_{\star}$ and $\beta$ (or $v_{\rm esc}$) are enough to rescale the equations in physical units.
Finally, the algebraic equation for $\Gamma$ can be obtained from equation \ref{eq_energy_int} and reads:
\begin{equation}\label{eq_Gamma}
\Gamma = \dot{\mathcal{E}} - \alpha \left( w + x - 1 + \frac{1+g(\tilde{\tau})}{1-x} \frac{u}{\tilde{\tau}} \right),
\end{equation}
where $\dot{\mathcal{E}} = \dot{E}/L_{\rm Edd}$ and $g(\tilde{\tau})$ is defined in equation \ref{eq_g_tau}.
The dependent variables $w$, $u$ and $\Gamma$ are proportional to the kinetic energy of the gas, to the energy density of the radiation and to the luminosity carried by photons, respectively.
Therefore, equations \ref{eq_w}, \ref{eq_u} and \ref{eq_Gamma} compose the system that describes the energy exchanges between the different components of the system.


\subsection{Numerical integration of the wind equations} \label{subsec_num_integr}

We integrate numerically the wind equations (equations \ref{eq_w}, \ref{eq_u}, \ref{eq_tau} and
\ref{eq_Gamma}) using the {\sc cvode} module of the {\sc sundials}\footnote{{\sc sundials} is publicly
  available at \url{https://computation.llnl.gov/casc/sundials/main.html}.} package
\citep{cohen+96,hindmarsh+05}.  {\sc cvode} is a C solver for stiff and non-stiff ordinary
differential equation systems in explicit form.  We adopt a fifth-order backward differentiation
formula in fixed-leading coefficient form with a modified Newton iteration to solve non-linear
systems.  {\sc cvode} provides also a module to find the roots of nonlinear equations which is well
suited to determine the position of the photosphere and the local properties of the wind while
contemporary solving the wind equations.

In order to find solutions of the wind equations, we follow the procedure outlined by
\citeauthor{quinn+85} (\citeyear{quinn+85}; see also \citealt{zytkow+72,kato+83}).  First of all, we
characterise our star by choosing a value for $\beta$.  Then, we pick a value for
$\alpha$ and $\dot{\mathcal{E}}$.  These two constants of integration are not enough to fully
characterise the wind.  We need a boundary condition, specifically the asymptotic gas
velocity at infinity $w_{\infty}$.  With that, we can specify the
initial conditions at a large radii $x_{\infty} = 1 - \delta$, much larger than $R_{\star}$ (i.e. when
$\delta \rightarrow 0$) and start our integration of the wind from outside inward.  Practically, we
calculate the radiative luminosity as seen by an observer at infinity as:
\begin{equation}\label{eq_lumin_infinity}
\mathcal{L}_{\infty} = \dot{\mathcal{E}} - \alpha w_{\infty}.
\end{equation}
At $x_{\infty}$,  the luminosity $\Gamma_{\infty}$ comoving with the flow can be considered constant and given by:
\begin{equation}\label{eq_Gamma_infty}
\Gamma_{\infty} = \frac{\mathcal{L}_{\infty}}{1 + 2 w_{\infty}^{1/2}/\beta} = 
\frac{\dot{\mathcal{E}} - \alpha w_{\infty}}{1 + 2 w_{\infty}^{1/2}/\beta}.
\end{equation}
We can then integrate the radiation energy density in the optically thin limit:
\begin{equation}
u' \simeq -2 (1-x) \Gamma_{\infty} \quad \Rightarrow \quad u = \Gamma_{\infty} (1-x)^2,
\label{eq:u_infty}
\end{equation}
where we use the boundary condition $u(1) = 0$, and the wind velocity:
\begin{equation}
w' = \Gamma_{\infty} - 1 \quad \Rightarrow \quad w = w_{\infty} -(\Gamma_{\infty} - 1) (1 - x).
\end{equation}
We can also write the explicit behaviour of $\tilde{\tau}$:
\begin{equation}
\tilde{\tau} = \frac{\alpha \beta (1-x)}{\sqrt{w_{\infty} -(\Gamma_{\infty} - 1) (1 - x)}}.
\label{eq:tau_infty}
\end{equation}
Note that equation \ref{eq_Gamma_infty} comes out naturally by evaluating equation \ref{eq_Gamma} at
$x \rightarrow 1$, where $1+g(\tilde{\tau}) \rightarrow 2$
and $u$ and $\tilde{\tau}$ are described by the expressions above (equations \ref{eq:u_infty} and \ref{eq:tau_infty}).
Next, we choose a value for
$\delta$, typically $\delta \sim 10^{-6}$, we check that indeed $\tilde{\tau}(x_{\infty}) \ll 1$ and
we use the formulas above to provide the initial conditions $w(x_{\infty})$ and $u(x_{\infty})$ for
the wind equations. 

We then integrate the equations inward up to the surface $x=0$ (or up to the point where a solution exists). 
We define the photosphere as the place where the equality
$L_{\rm phot} = 4 \pi R^2_{\rm phot} \sigma T_{\rm phot}^4$ is satisfied, where $\sigma = c a / 4$ is
the Stephan-Boltzmann constant, whereas the temperature $T$ is defined from the energy density $U$ as
$T = (U/a)^{1/4}$, regardless of the local optical depth.  Such a temperature is a proxy for the local
temperature and it recovers its full physical meaning only when $\tilde{\tau} > 1$.  The photosphere
identified in this way usually lays at $\tilde{\tau} \sim 2-3$.

Every solution of the wind equations is specified by the parameters $\alpha$, $\dot{\mathcal{E}}$ and
$w_{\infty}$, once the underlying star is set by $\beta$.  Among those parameters, $\alpha$ is
directly related to the outflow rate $\dot{M}$ and is necessary to solve the wind equations, i.e. such
a model does not allow to infer theoretically the value of $\dot{M}$ a priori.  However, an
\emph{acceptable} solution has to satisfy additional self-consistency requirements, which in turns
impose constraints of the parameter space and ultimately on the value of $\dot{M}$.  Those
self-consistency requirements are imposed by the assumption that the wind originates from a star-like object. In
particular: (i) the solution has to extend inward to at least $x=0$; (ii) the wind has to be optically
thick close to the surface of the star, i.e. the photosphere has to be above the base of the wind,
namely $\tilde{\tau}_{\star} > 1$ and $R_{\rm phot} > R_{\star}$; and (iii) the wind has to connect to
an hydrostatic solution, i.e. it should be initially subsonic (i.e. $\mathcal{M}_{\star} < 1$) 
and with a moderate
velocity\footnote{As consequence of the assumption of steady-state, we note that we cannot allow for
  $w_{\star} = 0$ because the density would otherwise diverge, as implied by the conservation of mass
  in equation \ref{eq_mass_cons}.} (i.e. $w_{\star} \ll 1$).  A wind solution is then accepted only
when it satisfies all the conditions listed above, and it is discarded otherwise.

\begin{figure}
\begin{center}
\includegraphics[width=8cm]{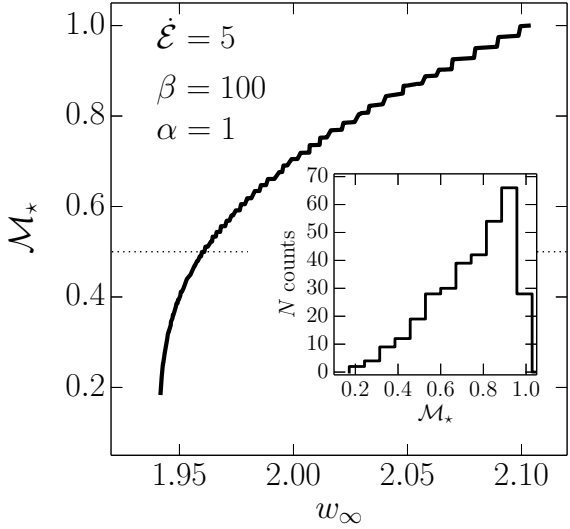}
\end{center}
\caption{Relation between the velocity at infinity $w_{\infty}$ and the Mach number at the surface
$\mathcal{M}_{\star}$ assuming $\beta=100$, $\alpha=1$ and $\dot{\mathcal{E}}=5$.  
The dotted thin line marks $\mathcal{M}_{\star} = 0.5$ for reference.
The range of possible $w_{\infty}$ is very narrow.
The inset shows the distribution of the Mach numbers $\mathcal{M}_{\star}$ obtained by 500 
realisations.
A clear peak around $0.8 - 0.9$ is present.
}
\label{fig_winf_mach}
\end{figure}


\subsubsection{The wind velocity at infinity} \label{subsec_mach_winfty}

To try and simplify further our procedure, we first asses the sensitivity of our solution to our
choice of $w_{\infty}$.  The arbitrariness of $w_{\infty}$ is simply a consequence of our neglecting
gas pressure, in the equations describing a radiation dominated wind.  When thermal gas pressure is
explicitly accounted for, the sonic point of a solution (i.e. where $\mathcal{M} = 1$) is also a
critical point (i.e. a divergent point for $w'$).
The requirement on the position of the critical
point to cure the local divergency translates naturally into a condition that fixes the value of
$w_{\infty}$.  As a consequence, solutions of the wind equations with gas pressure 
only dependent on $\alpha$ and
$\dot{\mathcal{E}}$ \citep[e.g.][]{quinn+85}.  In our solutions, instead, the sonic point is
\emph{not} a critical point, and $w_{\infty}$ is not univocally determined.  However, the fact that
with gas pressure terms there is only a single value for $w_{\infty}$ and that solutions should be
continuous as $p_{\rm gas}/p_{\rm rad} \rightarrow 0$ suggests that the range of possible $w_{\infty}$
may be narrow. Therefore, we investigate this possibility.

We setup a grid of  five representative values for $\beta \in \{10, 50, 100, 500, 1000 \}$, and five
representative values for $\dot{\mathcal{E}} \in \{ 1.5, 2.5, 5, 7.5, 10 \}$.  For each pair
$(\beta, \dot{\mathcal{E}})$, we divide the interval $\log \alpha \in [-3, 1]$ uniformly, and for each
value of $\alpha$ we run $10^3$ integrations of the wind equations choosing a random value for
$w_{\infty}$ distributed uniformly in the logarithmic interval $[-3, \log(\dot{\mathcal{E}}/\alpha)]$.
We keep only the acceptable solutions according to Section \ref{subsec_num_integr}.

Our results confirm that the range of $w_{\infty}$ that leads to self-consistent solutions is narrow,
usually $\lesssim 0.1$~dex, and centred around $w_{\infty} \sim 1$.  The values of $w_{\infty}$ also
correlate with $\mathcal{M}_{\star}$ in the interval $0.1 \lesssim \mathcal{M}_{\star} < 1$.  Such a
correlation is shown for 500 realisations in Figure \ref{fig_winf_mach} for an example configuration 
with $\beta=100$, $\alpha=1$ and $\dot{\mathcal{E}} = 5$ and exhibits typical features common to all the 
other combinations of parameters.  In particular, most of the interval of allowed $w_{\infty}$ corresponds
to values of the Mach number larger than $\sim 0.5 - 0.6$, as shown by the distribution of 
$\mathcal{M}_{\star}$ represented in the inset of Figure \ref{fig_winf_mach}, peaking around $\mathcal{M}_{\star} \sim 0.8$.  
Motivated by that, we can use this occurrence as an
approximate additional constraint to remove the freedom of choosing $w_{\infty}$ by choosing a
value for $\mathcal{M}_{\star}$ to be matched at $R_{\star}$. 
In the following, we focus only on solutions with $\mathcal{M}_{\star} = 0.8 \pm 0.05$.


\subsection{Results} \label{subsec:results_w}

\begin{figure}
\begin{center}
\includegraphics[width=8cm]{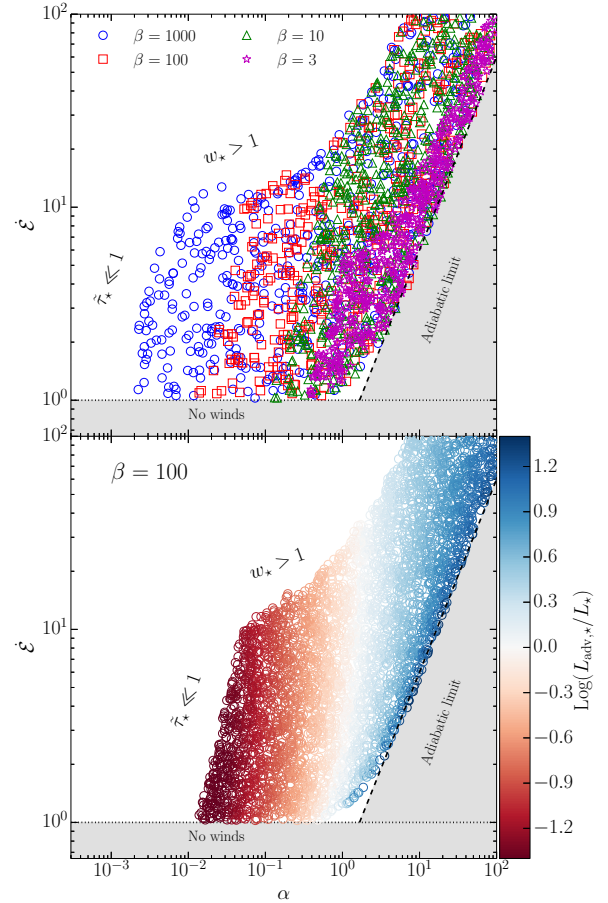}
\end{center}
\caption{Monte Carlo sampling of the parameter space $\alpha-\dot{\mathcal{E}}$.  Upper-panel: blue
  circles, red squares, green triangles and magenta stars show the results for $\beta = 1000$, 100,
  10, and 3, respectively.  For visualisation purposes, we show 10\% of the $10^4$ calculations
  performed for each value of $\beta$.  Lower panel: the same as above showing all the results for the
  case $\beta = 100$ where we colour-coded the points according to the ratio between the advected
  luminosity at the surface $L_{\rm adv,\star}$ and the radiative
  luminosity $L_{\star}$ at $R_{\star}$.  Both panels show the limits to the parameter space imposed
  by the self-consistency conditions as described in the text.  }
\label{fig_alpha_edot_plane}
\end{figure}

We are now in the position to explore the properties of the solutions within the parameter space
$(\alpha, \dot{\mathcal{E}}$) as a function of $\beta$.  Figure \ref{fig_alpha_edot_plane} highlights
the boundaries on the parameter space imposed by the self-consistency conditions. Solutions are
limited from below by the requirement that $\dot{\mathcal{E}} \geq 1$. We assume this condition as
necessary in order to launch the wind. In fact, the actual condition to have an accelerating wind is
$\Gamma_{\star} > 1$ or in other words, that the star should shine above the Eddington limit (equation
\ref {eq_w}). Nonetheless, we conveniently chose the limit $\dot{\mathcal{E}} \geq 1$ because (i) it
implies $\Gamma_{\star} > 1$ and (ii) the hydrostatic solution (the star) below the wind naturally
provides $\dot{E} \approx L_{\star}$ as a boundary condition (we discuss this point with more details
in Section \ref{sec_3}). The limit at small $\alpha$ and $\dot{\mathcal{E}} \lesssim 10$ results from
imposing $\tilde{\tau}_{\star} > 1$ and it depends on $\beta$ since the normalisation of the optical
depth is $\propto \alpha \beta$, as shown by equation \ref{eq_tau} and in the upper panel of
Figure \ref{fig_alpha_edot_plane}.  Physically, this is because matter needs to be faster to escape
from a more compact star, and from mass conservation ($\rho \propto \dot{M}/v$) it follows that a
higher mass loss rate is required to maintain the same optical depth $\tilde{\tau}_{\star} >
1$.
Finally, at fixed $\alpha$, the upper value of $\dot{\mathcal{E}}$ is constrained by the matching with
a hydrostatic solution below the wind, that requires $w_{\star} < 1$. Incidentally, we note here that
our set of prescriptions do not set an upper limit on $\dot{E}$.  This will be provided by the
physical characteristics of the stellar object powering the wind, more explicitly by how much super
Eddington its emission can be.  We will explicitly show this in Section \ref{sec_3} for quasi-stars.

\begin{figure*}
\begin{center}
\includegraphics[width=16cm]{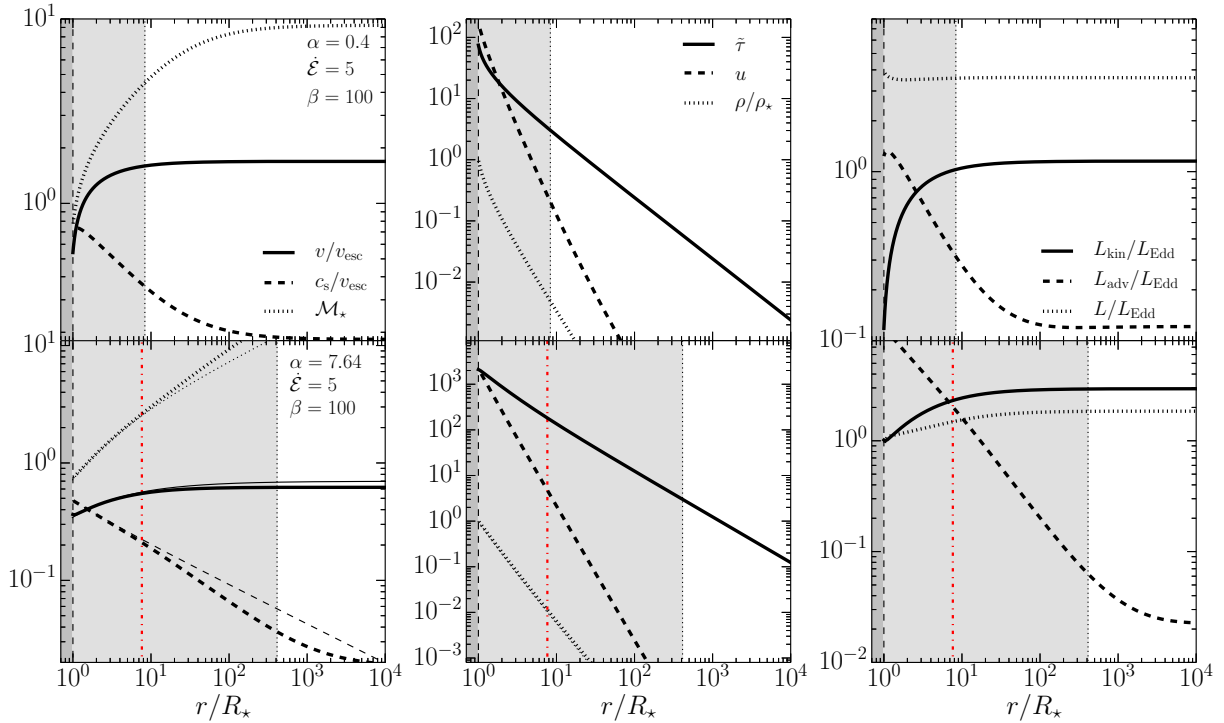}
\end{center}
\caption{Example solutions of the wind equations for $\beta = 100$.  The upper row shows the solution
  for $\alpha = 0.4$ and $\dot{\mathcal{E}} = 5$, while the bottom row shows the solution for
  $\alpha = 7.64$ and $\dot{\mathcal{E}}$, very close to the adiabatic limit of Figure
  \ref{fig_alpha_edot_plane}.  Left column: continuous, dashed and dotted lines show the profile of
  $v/v_{\rm esc}$, $c_{\rm s}/v_{\rm esc}$ and $\mathcal{M}$, respectively.  Central column:
  continuous, dashed and dotted lines show the profile of $\tilde{\tau}$, $u$ and $\rho/\rho_{\star}$,
  respectively.  Right column: continuous, dashed and dotted lines show the profile of
  $L_{\rm kin}/L_{\rm Edd}$, $L_{\rm adv}/L_{\rm Edd}$ and $L/L_{\rm Edd}$, respectively.  The dark
  and light grey shaded regions in all panels mark the base of the wind and the surface of the
  photosphere, respectively, while the red, vertical, dash-dotted lines in the bottom row indicate the
  position of the trapping radius.  The thin lines in the bottom-left panel and the thin dotted line
  in the bottom-central panel show the velocity, sound speed, Mach number and density, respectively,
  of the adiabatic wind with the same critical point of the wind solution.  The two solutions are very
  similar within $R_{\rm tr}$.}
\label{fig_wind_solution}
\end{figure*}

The maximum mass loss rate for a given $\dot{\mathcal{E}}$ (grey area on the right) is instead a
physical limit. This is obtained when the dominant energy source for the kinetic luminosity of the wind is
the enthalpy of the gas and $\dot{\mathcal{E}} \approx w_{\infty}~\alpha$. 
In practise, this is the behaviour of an adiabatic wind. We show this by considering a
 fully adiabatic solution. This latter has no radiative
luminosity in its governing equations (see Appendix \ref{sec_adab_review}), and the presence of a
critical point allows us to relate the velocity at infinity $w_{\infty, \rm adiab}$ to the condition
at the base of the wind. We can therefore derive that an adiabatic wind with
$\mathcal{M}_{\star} = 0.8$ will have $w_{\infty, \rm adiab} = 3 w_{\rm c} \approx 0.6$, where
$w_{\rm c}$ is the velocity at the critical point $w_{\rm c} = s_{\rm c}$, related to
$\mathcal{M}_{\star}$ by equation \ref{eq_ad_energy} and the relation plotted in Figure
\ref{fig_mach_adiab}. For a given $\dot{\mathcal{E}}$, we therefore have:
\begin{equation}
\alpha_{\rm max} \approx \dot{\mathcal{E}}/0.6.
\end{equation}
This relation is plotted as a dashed line in Figure \ref{fig_alpha_edot_plane} and clearly marks the
rightmost limit of our solutions and the beginning of the ``forbidden" region dubbed ``adiabatic
limit''.  

That our solutions tend towards an adiabatic behaviour as $\dot{M}$ increases, can be better
appreciated by looking at the lower panel of Figure \ref{fig_alpha_edot_plane}, which shows solutions for
$\beta = 100$, colour coded according to the ratio $L_{\rm adv,\star}/L_{\star}$, where
$L_{\rm adv} = \dot{M} (p + U) / \rho$ is the luminosity advected within the bulk outflow.  Across
$\alpha \sim 1$ the regime of the wind changes: for $\alpha <1$ energy in radiation is mainly
transported by diffusion while for $\alpha > 1$ advection becomes more and more dominant as the
outflow rate increases towards the adiabatic limit.  As mentioned before, this is exactly the physical
meaning of the trapping radius (see also Figure \ref{fig_wind_solution}, lower rightmost panel) and
$\alpha >1$ implies that the trapping radius occurs within the outflow, $R_{\rm tr} > R_{\star}$ (see
equation \ref{eq:rtr}).  Formally, a fully adiabatic solution has $R_{\rm tr} = \infty$ (i.e.
$\alpha = \infty$) and correspondently $L_{\rm adv,\star}/L_{\star} = \infty$. Advection is indeed the
only transport mechanism in an adiabatic wind.  We can go a step further and calculate $\dot{M}$ of a
solution \emph{relative} to the maximum possible mass loss rate.  The latter corresponds to that of an
adiabatic wind ($\dot{M}_{\rm ad}$, equation \ref{eq_mdot_adiabatic}) launched from the same star,
with the same initial conditions (i.e. the same $c_{\rm s, \star}$ and $\rho_{\star}$ at $R_{\star}$)
at the same $\dot{\mathcal{E}}$ (which for an adiabatic wind reads
$\dot{\mathcal{E}} = \dot{M}_{\rm ad} (v^2_{\star}/2 + 3 c_{\rm s,\star}^2 - G M_{\star}/R_{\star}) /
L_{\rm Edd}$):
\begin{equation}
\frac{\dot{M}}{\dot{M}_{\rm ad}} \approx  \left( 1 - \frac{L_{\star}}{\dot{E}} \right).
\end{equation}
When the contribution of the radiative luminosity $L_{\star}$ at $R_{\star}$ to the total energetic
budget becomes negligible, $\dot{M}$ approaches the adiabatic value.

The upper panel of Figure \ref{fig_alpha_edot_plane} shows the comparison of $10^4$ realisations for
four different values of $\beta$.  We find that the winds originating from more compact stars
(i.e. those with smaller $\beta$) sustain higher outflow rates at the same total luminosity
$\dot{\mathcal{E}}$.  This is again related to the fact that gas needs to be faster to escape from a
more compact star, and it compensates this increase in velocity by an increase in $\dot{M}$ to
maintain optical thickness. As a consequence, an \emph{optically-thick} wind needs to carry a larger
and larger fraction of the total luminosity in kinetic and advective form. This explains why the range
of possible $\dot{M}$ gets squeezed towards the adiabatic limit for\footnote{Formally, such a limit
  cannot be reliably modelled by our wind equations, because it would imply approaching a regime where
  general relativistic corrections might become relevant.} $\beta \rightarrow 1$.

Figure \ref{fig_wind_solution} shows two example solutions of the wind equations with $\beta = 100$.
The upper row shows a solution with $\alpha = 0.4$ and $\dot{\mathcal{E}}=5$, i.e. at the centre of
the allowed region in the lower panel of Figure \ref{fig_alpha_edot_plane}, while the bottom row shows
a solution with $\alpha = 7.64$ and $\dot{\mathcal{E}}=5$, i.e. very close to the adiabatic limit.
The wind velocity of the first solution grows steeply within the photosphere and then flattens to its
asymptotic values; on the contrary, the sound speed $c_{\rm s}$ decreases quickly within the
photosphere, matching the gas velocity at the sonic point very close to $R_{\star}$, beyond which the
wind becomes highly supersonic.  Note that, regardless of the optical depth, we always use the general
definition of sound speed:
\begin{equation}\label{eq_sound_speed}
c_{\rm s}^2 =\left( \pderiv{p}{\rho} \right)_{\mathcal{S}} = \frac{4}{3} \frac{p}{\rho},
\end{equation}
where $\mathcal{S}$ is the specific entropy and the second equality is based on the most general
equation of state for a non-isentropic, radiation-pressure dominated fluid, namely
$p(\rho, \mathcal{S}) = K(\mathcal{S}) \rho^{4/3}$.  In fact, this relation is only valid when
$p = U / 3$, i.e. in the optically-thick limit, but we need to consider the behaviour of $c_{\rm s}$
only within the photosphere.  The solution close to the adiabatic limit shows a similar behaviour,
though the gas velocity grows less steeply than in the previous case and mostly within $R_{\rm tr}$.
It also reaches an asymptotic velocity lower than the previous case, though the total energy
$\dot{\mathcal{E}}$ is the same.

The profiles of the optical depth, density and radiation energy density are similar in both examples.
We can fairly accurately describe them with power laws, at least close to and outside the photosphere.
In particular, the optical depth decreases with radius as $r^{-1}$, while the density as $r^{-2}$ once
the gas velocity remains almost constant.  This confirms a posteriori that $\tilde{\tau}$ is a good
approximation for $\tau$.  We note also that the wind is highly optically thick close to $R_{\star}$,
while the photosphere forms at $\tilde{\tau}_{\rm phot} \sim 2-3$.  This latter confirms that our
approximate treatment of the transition between the optically thin and thick regimes gives sensible
results, close to the conventional $\tau_{\rm phot} \approx 1$.

The largest difference between the two solutions is in the energy budget, shown in the left column of
Figure \ref{fig_wind_solution}.  There, we compare the profiles of the radiative luminosity $L$, the advected
luminosity $L_{\rm adv}$, and the kinetic luminosity $L_{\rm kin} = \dot{M} v^2 / 2$, all normalised
to $L_{\rm Edd}$ for convenience.  These luminosities, summed up with $-G \dot{M} M_{\star} /r$, give
the total, constant luminosity $\dot{E}$.  The energy budget in the first solution is dominated by the 
radiative luminosity from the base of the wind (indeed $R_{\rm tr} < R_{\star}$) to infinity. 
After a small decrease, it remains almost constant with radius, while $L_{\rm kin}$ quickly rises within the photosphere, yet remaining
subdominant. This behaviour indicates that the acceleration of the wind is not powered by $L$ but it rather  
 occurs at the expenses of $L_{\rm adv}$, that drops accordingly within the photosphere.  
In contrast, $L_{\rm adv}$ in the $\alpha=7.64$ case (lower panel) is
initially an order of magnitude higher than both $L$ and $L_{\rm kin}$ and becomes comparable to $L$
around $R_{\rm tr}$.  Instead, $L$ remains always subdominant compared to $L_{\rm kin}$. 
The radiative luminosity's mild growth within the photosphere is similar to that of an ``effective'' Eddington ratio that one would obtain from equation
\ref{eq_w}, $\Gamma_{\rm eff} = 1 + w'$, when $w'$ from the adiabatic equations \ref{eq_ad_w} and
\ref{eq_ad_en} is used.  Notably, also in this case, $L_{\rm kin}$ is ultimately powered by
$L_{\rm adv}$.

Finally, we show explicitly the similarity between the solution close to the adiabatic limit and an
actual adiabatic solution in the bottom row of Figure \ref{fig_wind_solution}.  Specifically, after
calculating the solution of our wind equations, we calculate also the adiabatic solution (see Appendix
\ref{sec_adab_review}) which has the same critical radius.  Such an adiabatic solution crosses
naturally the critical point with the same value of critical velocity $w_{\rm c}$ of the full
solution, and closely resembles it (comparing $v$, $c_{\rm s}$, $\mathcal{M}$, and $\rho$) within the
trapping radius, outside of which radiative diffusion becomes relevant.

From our results, it is clear that the ultimate source of kinetic energy for the gas is advection energy and \emph{not} the diffusive radiation luminosity. 
This is not consumed as the gas propagates outward. 
To increase the mass loss, it is therefore necessary to go towards an adiabatic solution, where initially the advection energy dominates the energy budget. 
This is different from the results by \citet{shaviv+01b}, \citet{owocki+04}, and from the description used in \citet{dotan+11} of a ``photon-tired'' wind.


\section{Winds from Quasi-stars} \label{sec_3}

Directly forming massive black hole seeds is possible in principle when a mass of
$\sim 10^8$~M$_{\sun}$ can be collected in $\sim 10^6$~yr, requiring inflow 
rate $> 100$~M$_{\sun}$~yr$^{-1}$. 
The reason is that such a rapid accumulation of mass has to occur before
nuclear reactions dominate the evolution, setting a lifetime of $\sim 10^6$~yr \citep{begelman+10}.
In these extreme conditions, a dynamical instability due to
relativistic effects may develop and even rotation cannot prevent the cloud from 
collapsing directly into a black hole (\citealt{fowler+66,baumgarte+99,shibata+02}; see however \citealt{ferrara+14}).
On the other hand, milder conditions  (e.g. inflow rate $\sim 0.1-1$~M$_{\sun}$~yr$^{-1}$) 
would lead to the formation of a supermassive star, possibly $\sim 10^{5}-10^{7}$~M$_{\sun}$, 
stabilised by some rotation \citep{begelman+10,hosokawa+12,hosokawa+13}.
After $\sim 10^6$~yr, a small embryo seed ($\lesssim 100$~M$_{\sun}$) can form
at the centre of a supermassive star at the end of the hydrogen-burning phase \citep{begelman+10}.
Such a seed then needs to go through a
phase of vigorous super-Eddington accretion to reach $10^4-10^5$~M$_{\sun}$ within a few million years
from its birth.  This super-Eddington accretion can occur within a quasi-star: a very massive
($>10^5-10^6$~M$_{\sun}$) quasi hydrostatic envelope that surrounds the black hole and feeds it
at a rate equal to roughly its own (i.e. quasi-star's) Eddington limit \citep{begelman+08}.
Accretion around the Eddington limit involves radiation dominated gas, which is loosely bound with a
total energy close to zero.  For this reason outflows can easily form.  Here we investigate whether
this vigorous accretion in quasi-stars is also accompanied by mass loss, as expected in other
super-Eddington systems such as discs \citep{blandford+04}.


\subsection{Equations} \label{subsec_equation_qs}

We follow \citet{begelman+08} and \citet{dotan+11} to describe the hydrostatic envelope of a quasi-star and then we match it with our wind model, looking for equilibrium solutions.
A quasi-star is made of four components: (i) the central, accreting black hole, (ii) a convective, radiation-pressure dominated envelope, (iii) a porous radiative layer, and (iv) a wind.
In the following, we briefly describe components i-iii, remanding to \citet{dotan+11} for additional details.
We do not model explicitly the inflow on to the central black hole; instead, we treat it as a boundary condition, assuming that a black hole of mass $M_{\bullet}$ is accreting through a convection-dominated disc \citep{stone+99,igumenshchev+99,quataert+00,agol+01} within a few Bondi radii $r_{\rm B} = G M_{\bullet}/(2 c_{\rm s,c}^2)$, where $c_{\rm s,c}$ is the central sound speed $\lesssim r_{\rm B}$.
We assume that the black hole is radiating at a luminosity $L_{\bullet}$ close to the maximum that convection-dominated accretion flows can sustain evaluated at $5 r_{\rm B}$, namely:
\begin{equation}\label{eq_L_bh}
L_{\bullet} = L_{\rm conv}(5 r_{\rm B}) = 4~\pi~(5 r_{\rm B})^2~\rho_{\rm c}~c_{\rm s,c}^3, 
\end{equation}
where $\rho_{\rm c}$ is the central density outside a few $r_{\rm B}$.
$L_{\bullet}$ it is injected at the centre of the envelope and transported till the base of the wind, first by convection and then by diffusion.
We used it as an inner boundary condition for the integration of the envelope and we also neglect the gas mass within $5 r_{\rm B}$, assuming that only $M_{\bullet}$ contributes at smaller radii.
We checked the effect of varying the position of the inner boundary.
We find that most of the properties in the $M_{\bullet} - M_{\star}$ plane (see Section \ref{subsec:results_qs}) remain unchanged.
However, as discussed by \citet{ball+11} and \citet{ball+12}, changing the inner radius modifies (at the same central pressure) the ratio $M_{\bullet}/M_{\star}$ of two consistent solutions.
This slightly displaces the no-hydrostatic-solution region.
Although a few Bondi radii are a reasonable estimate for the inner accretion region where the black hole gravity is expected to dominate, we caution that such a choice remain somewhat arbitrary.

Outside $5 r_{\rm b}$, the radiation-pressure dominated, convective envelope extends; we assume that it satisfies the hydrostatic equilibrium:
\begin{equation}\label{eq_hydro_eq}
\deriv{P}{r} = - \frac{G M(r) \rho}{r^2},
\end{equation}
where the total pressure $P$ is the sum of the gas pressure $P_{\rm g} = \rho k_{\rm B} T / (\mu m_{\rm p})$ and the radiation pressure $P_{\rm r} = a T^4/3$, specified by the gas density $\rho$ and temperature $T$.
$k_{\rm B}$, $a$ and $m_{\rm p}$ are the Boltzmann constant, the radiation constant and the mass of the proton, respectively.
We assume the mean molecular weight $\mu = 0.59$, appropriate for gas with primordial composition at $T > 10^4$~K, as usually true everywhere in the interior of quasi-stars.
The enclosed mass $M(r)$ is given by:
\begin{equation}\label{eq_mass_enc}
M(r) = M_{\bullet} + 4 \pi \int_{5 r_{\rm B}}^{r} \rho(r') {r'}^{2} \de r'.
\end{equation}
The dominant energy transport mechanism within the envelope is convection, which induces a temperature gradient very close to adiabatic; therefore, we evolve the temperature gradient assuming that it is equal to the adiabatic one:
\begin{equation}\label{eq_adiab_grad}
\deriv{\log T}{\log P} = \frac{\gamma_{\rm ad} - 1}{\gamma_{\rm ad}},
\end{equation}
where the adiabatic index $\gamma_{\rm ad}$ depends on the ratio $\zeta \equiv P_{\rm g}/P$ according to:
\begin{equation}
\gamma_{\rm ad} = \frac{32 - 24 \zeta - 3 \zeta^2}{24 - 18 \zeta - 3 \zeta^2}.
\end{equation}

The convective envelope extends till the radius $r_{\rm conv}$ where $L_{\rm conv}(r_{\rm con}) = L_{\bullet}$.
Outside $r_{\rm conv}$, $L_{\rm conv} < L_{\bullet}$ and diffusion becomes more efficient in transporting $L_{\bullet}$ than convection.
However, $L_{\bullet}$ may be larger than the local Eddington limit associated to the local enclosed mass $M(r)$ and to electron scattering opacity.
In such a condition, the gas becomes locally unstable and develops inhomogeneities \citep{shaviv+01a} that have the effect of reducing the effective opacity with respect to its microscopic value even before than the Eddington limit is reached.
Following \citet{dotan+11}, we model the effective opacity $\kappa_{\rm eff}$ as:
\begin{equation}
\kappa_{\rm eff} = 
\left\{
\begin{array}{lc}
\displaystyle \frac{\kappa}{\Gamma} \left(1 - \frac{0.16}{\Gamma} \right) & \Gamma > 0.8, \\
\kappa & \Gamma \leq 0.8,\\
\end{array}
\right.
\end{equation}
where $\Gamma = L_{\bullet} / L_{\rm Edd}$ is the local Eddington ratio calculated using equation \ref{eq_eddington} with $M(r)$, while $\kappa$ is the microscopic opacity:
\begin{equation}\label{eq_opacity_temp}
\kappa(T) = \frac{\kappa_{\rm es}}{1 + (T/T_0)^{-13}},
\end{equation}
where $T_0 = 8000$~K.
This opacity models the results for pristine gas by \citet{mayer+05}.
The effective opacity corresponds to a an effective Eddington ratio $\Gamma_{\rm eff} = 1 - 0.16/\Gamma$ when $\Gamma > 0.8$, i.e. the gas is effectively sub-Eddington, though it would be super-Eddington with the microscopic opacity.
Throughout this radiative layer (that usually encompasses a tiny fraction of the total mass), we assume once again hydrostatic equilibrium (being effectively sub-Eddington) and we solve equations \ref{eq_hydro_eq} and \ref{eq_mass_enc}, but we evolve the temperature by mean of the radiative gradient with $\kappa_{\rm eff}$:
\begin{equation}\label{eq_rad_grad}
\deriv{T}{r} = - \frac{3 \kappa_{\rm eff} \rho L_{\bullet}}{16 \pi a c r^2 T^3}.
\end{equation}
The luminosity $L_{\bullet}$ remains constant since no energy sources/sinks are present within the convective envelope or the radiative layer.
The inhomogeneities in the radiative layer can maintain the luminosity sub-Eddington as long as they remain optically thick.
Since those inhomogeneity have a size of order of the local density scale-height, we can estimate their optical depth as $\tau_{\rm eff} \approx \chi \rho \kappa_{\rm eff} h$, where $\chi$ is the ionisation fraction calculated from the Saha equation assuming equilibrium, and $h = |\rho / (\de \rho / \de r)|$ is the density scale-height.
Then, the radiative layer extends up to $r_{\rm rad}$ such that $\tau_{\rm eff}(r_{\rm rad}) = 1$.
We note that $\zeta$ typically decreases quickly throughout the radiative layer, reaching values $\zeta \ll 0.01$ at $r_{\rm rad}$.

We can finally solve and connect the wind model described in Section \ref{sec_2}.
In particular, we use $M_{\star} = M(r_{\rm rad})$ and $R_{\star} = r_{\rm rad}$.
From these quantities we can evaluate $\beta$ associated to the star.
As in Section \ref{sec_2}, we do not model the initial acceleration of mass explicitly.
Instead, we assume that this occurs very quickly around $r_{\rm rad}$, which represent the interface between the hydrostatic part and the wind.
Then, we assign $\dot{E} = L_{\bullet}$ to guarantee the conservation of energy at the interface, because below $r_{\rm rad}$ there is no net displacement of mass and the total luminosity transported is just $L_{\bullet}$.
Finally, we need to specify $\dot{M}_{\rm wind}$.
As described in Section \ref{subsec:results_w}, we assume that the wind connect with the hydrostatic part with a fixed $\mathcal{M}_{\star} < 1$.
At the same time, we assume continuity for the density and pressure at $r_{\rm rad}$, which implies:
\begin{equation}
\dot{M}_{\rm wind} = 4~\pi~r_{\rm rad}^2 \rho_{\star} \mathcal{M}_{\star} c_{\star},
\end{equation}
where $\rho_{\star} = \rho(r_{\rm rad})$ and $c_{\star} = c_{\rm s}(r_{\rm rad})$ are the density and the sound speed evaluated at $r_{\rm rad}$ as given by the integration of the radiative layer, respectively.
Once we have $\beta$, $\dot{E}$ and $\dot{M}$, we can integrate the wind equations as described in Section \ref{subsec_num_integr} and \ref{subsec:results_w}, with the only difference that we use the temperature-dependent opacity of equation \ref{eq_opacity_temp}.
We evaluate it using as a proxy for the local temperature $T = (U/a)^{1/4}$, which is correct only in the optically-thick part of the atmosphere.
We discuss the limitations of these assumptions in Section \ref{sec_5}.


\subsection{Numerical integration}\label{subsec_num_integr_qs}

\begin{figure*}
\begin{center}
\includegraphics[width=16cm]{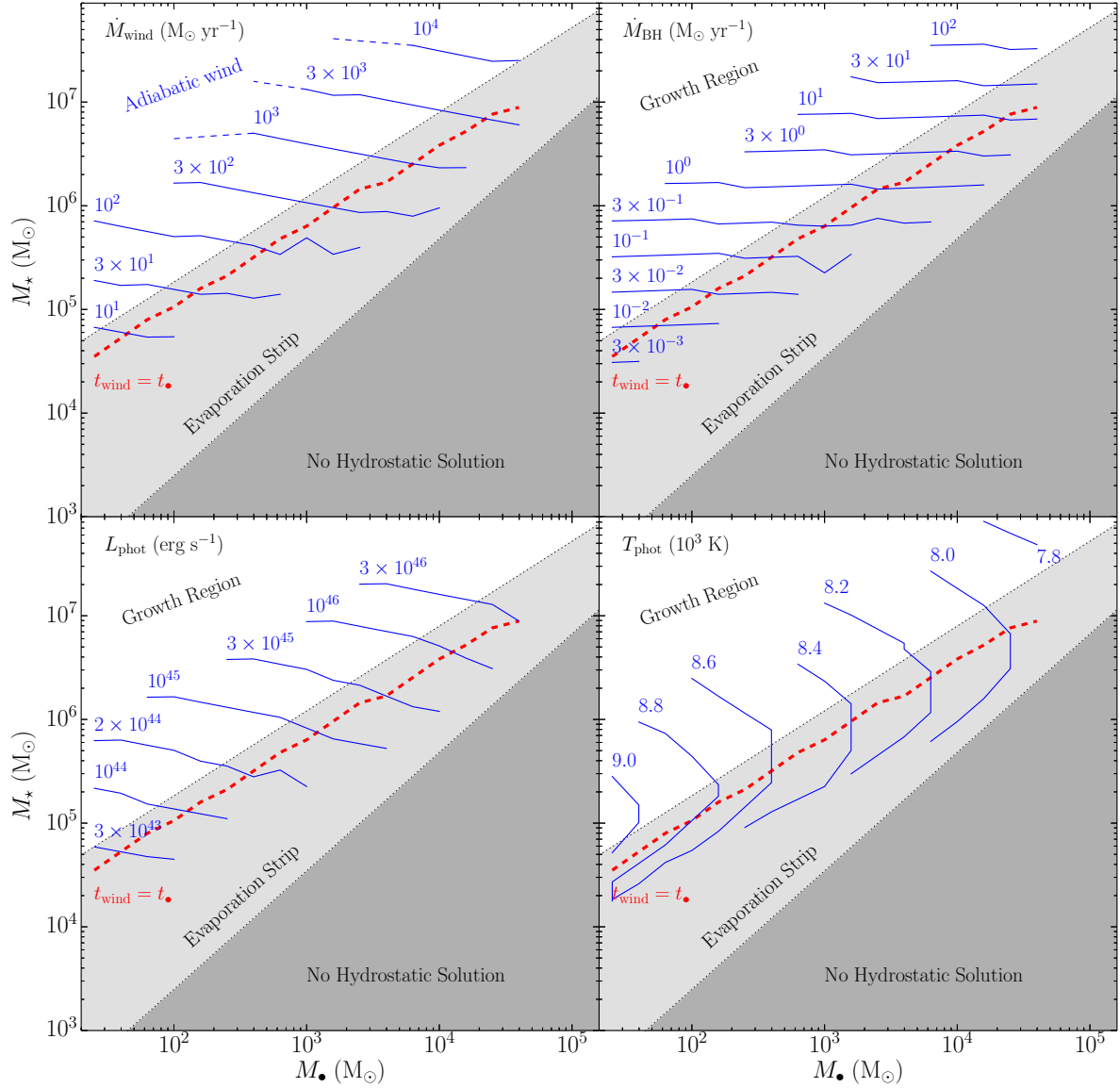}
\end{center}
\caption{Properties of our quasi-star models in the $M_{\bullet}-M_{\star}$ plane.
From upper-left panel, clockwise:  thin, blue continuous lines show the isocontours of the outflow rate $\dot{M}_{\rm wind}$, of the accretion rate on to the black hole $\dot{M}_{\bullet}$, of the photospheric bolometric luminosity $L_{\rm phot}$, and of the photospheric effective temperature $T_{\rm phot}$, respectively.
The blue, dashed lines in the upper-left panel show the extension in the adiabatic wind regime of some isocontours as discuss in the text.
The adiabatic wind region is labelled explicitly in this panel, while the label ``Growth Region'' is omitted for clarity.
In each panel, the white, light grey and dark grey shaded regions represent the growth region, the evaporation strip (i.e. where $t_{\rm wind} < t_{\bullet}$), and the region where no hydrostatic solutions exist as reported by \citet{dotan+11}, respectively; on the other hand, the thick, red, dashed line shows the limit of the evaporation strip (i.e. where $t_{\rm wind} = t_{\bullet}$) for our models.
}
\label{fig_mbh_ms_plane}
\end{figure*}

We proceed to describe the numerical strategy to solve the equations in Section \ref{subsec_equation_qs}, similar to what has been done by \citet{dotan+11}.
\begin{enumerate}
\item We choose the black hole mass $M_{\bullet}$.
\item We choose one of the central quantity, in particular the total central pressure $P_{\rm c}$.
\item We need a second quantity to specify all the boundary conditions at the centre.
Therefore, we guess the value of $\zeta_{\rm c} = P_{\rm gas, c}/P_{\rm c}$.
\item We calculate the central quantities: $T_{\rm c} = 3 (1 - \zeta_{\rm c}) P_{\rm c} / a$, $\rho_{\rm c} = P_{\rm c} \zeta_{\rm c} \mu m_{\rm p} / (k_{\rm B} T_{\rm c})$, $c_{\rm s,c}^2 = ((4/3) (1 - \zeta_{\rm c}) + (5/3) \zeta_{\rm c}) P_{\rm c} / \rho_{\rm c}$ and $r_{\rm B} = G M_{\bullet} / (2 c_{\rm s,c}^2)$; we evaluate $L_{\bullet}$ according to equation \ref{eq_L_bh}.
\item We integrate the convective envelope, namely equations \ref{eq_hydro_eq}, \ref{eq_mass_enc} and \ref{eq_adiab_grad}, from the centre ($5 r_{\rm B}$) outward, until we reach $r_{\rm conv}$.
We actually integrate the equation in their Lagrangian form, using the enclosed mass $M(r)$ as the independent variable.
\item We integrate the radiative layer equations \ref{eq_hydro_eq}, \ref{eq_mass_enc} and \ref{eq_rad_grad}, assuming continuity with the convective envelope from $r_{\rm conv}$ till $r_{\rm rad}$.
We integrate the equations using $P$ as the independent variable since it varies more than the other quantities throughout the radiative layer.
\item We calculate the necessary quantities to specify the properties of the wind using the values at $r_{\rm rad}$ as discussed in Section \ref{subsec_equation_qs}, namely $\beta$, $\dot{E}$ and $\dot{M}_{\rm wind}$; specifically, we assume $\mathcal{M}_{\star} = 0.8$.
\item We calculate the wind solution and we check that/whether it reaches $r_{\rm rad}$ self-consistently as described in Section \ref{subsec_num_integr}.
\item We compare the density $\rho^{\rm (rad)}(r_{\rm rad})$ obtained at the end of the integration of the radiative layer with the density $\rho^{\rm (wind)}(r_{\rm rad})$ obtained by the wind integration and we modify $\zeta_{\rm c}$ in order to match the two values.
\end{enumerate}

We find empirically that the ratio $\rho^{\rm (wind)}(r_{\rm rad})/\rho^{\rm (rad)}(r_{\rm rad})$ crosses the value 1 extremely steeply while varying $\zeta_{\rm c}$ and is not monotone far from the solution. 
This occurrence makes difficult to use classic methods such as bisection unless the initial guesses for the values of $\zeta_{\rm c}$ that bracket the final solution are very close to the latter.
To overcome this problem, we proceed as follow: we choose an initial guess for $\zeta_{\rm c}$ by solving the following equation that comes from the scaling relations of the envelope provided by \citet{begelman+08} and \citet{dotan+11}:
\begin{equation}
p_{\rm c,7} = \frac{1.134}{m_{\bullet}^{16/5}} \frac{(1-\zeta_{\rm c})^{2/5}}{\zeta_{\rm c}^{28/5}},
\end{equation}
where $P_{\rm c} = p_{\rm c, 7} \times 10^7$~erg~cm$^{-3}$ and $M_{\bullet} = m_{\bullet}$~M$_{\sun}$.
Then, we build a grid of models for several values of $\zeta_{\rm c}$ around the initial guess, and we progressively refine this grid around the solution.
When we bracket the true solution with a relative precision $\sim 10^{-3}$, we use this bracketing as the starting points for a Brent root-finder \citep{brent+73,press+02}.
The typical solutions of $\zeta_{\rm c}$ are $\lesssim 0.01$; indeed, quasi-stars are radiation-pressure dominated in their interiors.
Moreover, we note that $\zeta$ typically decreases throughout the radiative layer, reaching values $\zeta \ll 0.01$ at 
$r_{\rm rad}$.
This behaviour justifies our simplifying assumption of neglecting the gas pressure in the wind, since the ratio $3 k_{\rm B} \rho / (\mu m_{\rm p} a T^3)$ remains effectively $\ll 1$ through the wind and within the photosphere. Calculations a posteriori of $\zeta$ in the wind show a decreasing behaviour. This suggests that our treatment is at least consistent. Of course, the radiation dominated assumption limits our results to very massive stars and cannot be extended to e.g. Population III stars.


\subsection{Results for quasi-stars} \label{subsec:results_qs}

We run a grid of models exploring a wide range of $M_{\bullet}$ and $P_{\rm c}$, which maps into $M_{\star}$.
Figure \ref{fig_mbh_ms_plane} summarises our findings in the $M_{\bullet}-M_{\star}$ plane.
Specifically, the various panel shows iso-contours of the outflow rate $\dot{M}_{\rm wind}$, the accretion onto the black hole $\dot{M}_{\rm BH} = L_{\bullet}/(\eta c^2)$, where the radiative efficiency $\eta \simeq 0.1$, the photospheric luminosity $L_{\rm phot}$, and the effective, photospheric temperature $T_{\rm phot}$.

Such a plane is characterised by three regions. 
For high black hole masses and relatively low envelope masses, no hydrostatic solution can be found \cite[see also ][]{begelman+08}.
That is because the quasi-stars would stay beyond the \citet{hayashi+61} track, which represent a lower limit to the effective temperature (around 4000~K) of a convective envelope in hydrostatic equilibrium.
Beyond such a limit, no solutions for the hydrostatic envelope exist.
The second region is the evaporation strip identified by \citet{dotan+11}.
This region lays where the evaporation timescale $t_{\rm wind} = M_{\star} / \dot{M}_{\rm wind}$, i.e. the typical timescale over which the stellar envelope would be blown away by the winds, is shorter than the accretion timescale $t_{\bullet} = M_{\bullet} / \dot{M}_{\rm BH}$.
Within this region, the envelope mass is dispersed by the wind before than the black holes can accrete further.
The third region is the growth region, where the black hole can accrete substantial mass from the envelope ($t_{\bullet} < t_{\rm wind}$).

\begin{figure}
\begin{center}
\includegraphics[width=8cm]{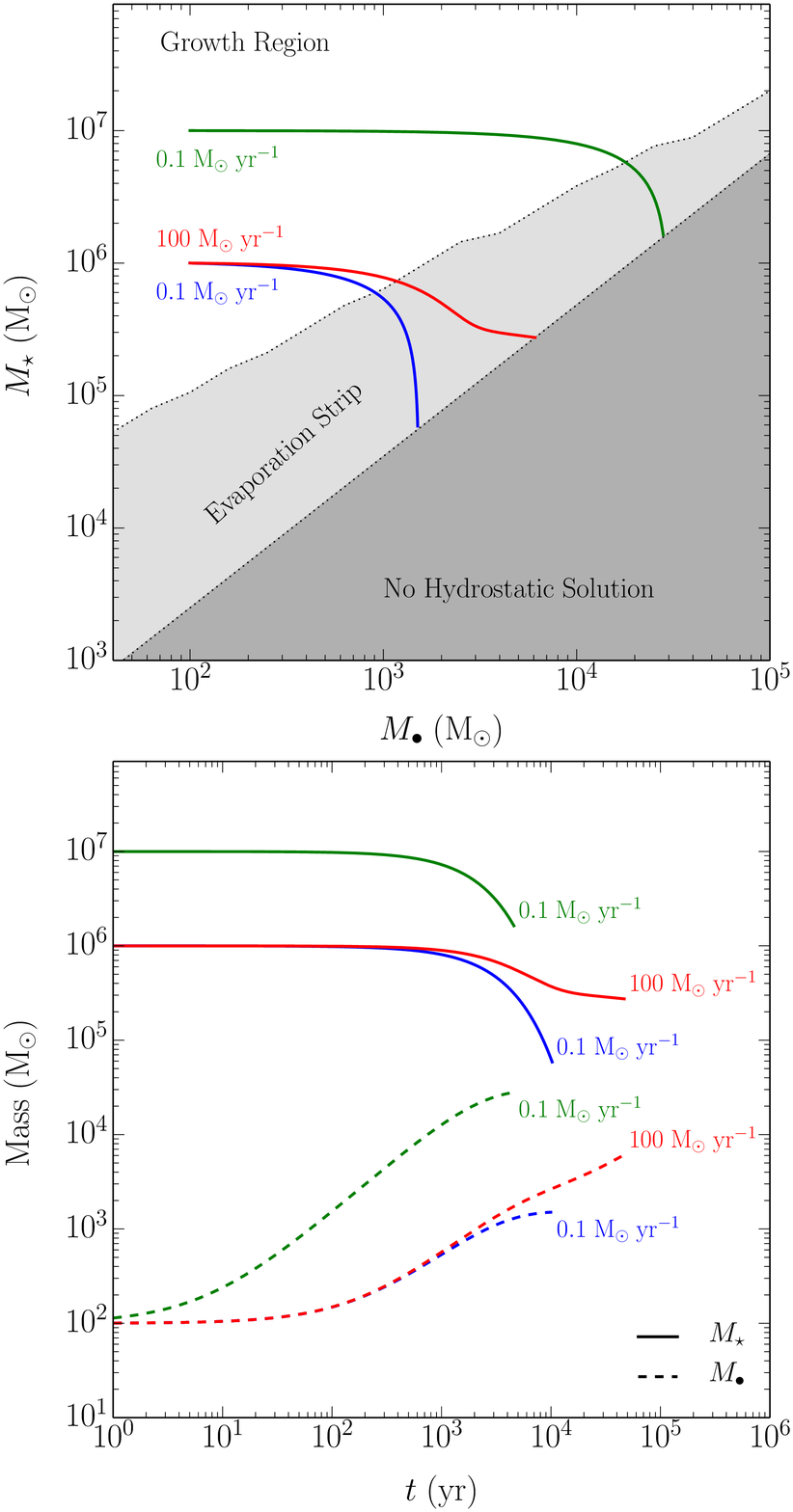}
\end{center}
\caption{
Evolutionary tracks for quasi-stars.
Upper panel: evolutionary tracks in the $M_{\bullet}-M_{\star}$ plane (see Figure \ref{fig_mbh_ms_plane} for a description, but note that here we plot the evaporation strip determined from our models).
The green line corresponds to a quasi-stars with initial $M_{\bullet}= 100$~M$_{\sun}$, $M_{\star}=10^{7}$~M$_{\sun}$, and $\dot{M}_{\rm in} = 0.1$~M$_{\sun}$~yr$^{-1}$.
The blue and red lines both correspond to a quasi-star with initial $M_{\bullet}= 100$~M$_{\sun}$, $M_{\star}=10^{6}$~M$_{\sun}$, but with $\dot{M}_{\rm in} = 0.1$~M$_{\sun}$~yr$^{-1}$ and $\dot{M}_{\rm in} = 100$~M$_{\sun}$~yr$^{-1}$, respectively.
Lower panel: evolution of $M_{\bullet}$ and $M_{\star}$ as a function of time.
Continuous and dashed lines (coupled with colours) show the evolution of $M_{\star}$ and $M_{\bullet}$, respectively.
}
\label{fig_qs_evol}
\end{figure}

We find that the iso-contours of  $\dot{M}_{\rm wind}$ due to our wind model of Section \ref{sec_2} are different from those in \citet{dotan+11}.
Most notably, we do not see any effect of photon-tiring, consistently with our result of Section \ref{subsec:results_w} that the ultimate source of the wind kinetic luminosity is the advection luminosity due to internal energy.
Specifically, our contours increase mildly when $M_{\bullet}$ decreases, while their would become suddenly much more steep at the onset of photon-tiring, with $\dot{M}_{\rm wind}$ almost independent of $M_{\star}$.
At constant $M_{\bullet}$, the different shapes of the iso-contours are such that $\dot{M}_{\rm wind}$ grows faster at high values of $M_{\star}$ until it reaches the adiabatic limit discussed in Section \ref{subsec:results_w}.

When our models hit the adiabatic limit, our quasi-star structures are all characterised by
$s_{\star} = c_{\rm s, \star}^2 / v_{\rm esc}^2$ very close to the value associated to $\mathcal{M}_{\star} \simeq 0.8$ for an
adiabatic wind, as shown by Figure \ref{fig_mach_adiab} in the Appendix \ref{sec_adab_review}.
Above this limit, our choice of a specific surface Mach number ($ \approx 0.8$) over-constrains our mathematical system and solutions cannot be found.
Technically, we would need to leave $\mathcal{M}_{\star}$ free to vary and the likely result would be solutions with nearly adiabatic winds.

To test this expectation,   we then change method and consider the idealised case of a purely adiabatic wind solution (see Appendix \ref{sec_adab_review}) and match it to the hydrostatic envelope. These winds are completely specified by the values of the density $\rho(r_{\rm rad})$ and of the sound speed $c_{\rm s}(r_{\rm rad})$ 
at the end of the envelope. We then check that $s$ associated to $c_{\rm s}(r_{\rm rad})$ is in the interval $1/6 < s < 1/4$ in order to have a solution with
$\mathcal{M}_{\star} < 1$ (Figure \ref{fig_mach_adiab}).
Finally, we calculate the adiabatic mass loss $\dot{M}_{\rm ad}$ according to equation \ref{eq_mdot_adiabatic} and we choose the equilibrium
solution (i.e. the value of $\zeta_{\rm c}$) that allows us to match the luminosity $L_{\bullet}$ to the luminosity carried by the adiabatic wind
evaluated at $R_{\star} = r_{\rm rad}$, namely
$3 \dot{M}_{\rm ad} c_{\rm s,\star}^2 (1 + \mathcal{M}^2_{\rm ad}(c_{\rm s, \star}) - v_{\rm esc}^2/(6 c_{\rm s, \star}^2) )$.
We find a fairly smooth transition between the two kinds of models, confirming the nearly adiabatic nature of the expected wind when this limit
is exceeded (see dashed lines in Figure \ref{fig_mbh_ms_plane}, top left panel).

This difference has also the effect of displacing slightly the interface between the evaporation strip and the growth region, i.e. the ``threshold-growth line'' \citep{dotan+11}.
Our models predict that such a line moves by a factor of $\sim 2- 3$ toward higher $M_{\bullet}$, decreasing the thickness of the evaporation strip.
Moreover, for every point $(M_{\bullet}, M_{\star})$ around and within the evaporation strip, we find an $\dot{M}_{\rm wind}$ smaller by a factor $\lesssim 10$ compared to \citet{dotan+11}, which implies a less sudden removal of the envelope when the quasi-stars enters the evaporation strip.
On the other hand, the iso-contours of $\dot{M}_{\rm BH}$ are almost independent of $M_{\bullet}$ and they are similar to previous findings.

The absence of a photon-tired wind has a strong impact on the photospheric luminosity of the quasi-stars.
Since the wind is mostly accelerated at the expense of the internal energy, the diffusive luminosity coming out at the photosphere $L_{\rm phot}$ is a large fraction of the luminosity $L_{\bullet}$ originally produced by central accretion and transported through the hydrostatic envelope.
Such a luminosity corresponds to Eddington ratios calculated with respect of the whole envelope mass that range from $\Gamma_{\rm phot} \sim 1$ to $\Gamma_{\rm phot} \lesssim 10$, even in the growth region, where \citet{dotan+11} find a decrease of $\Gamma_{\rm phot}$ due to photon-tiring.
Our models predict photospheric luminosities in the interval $10^{43} \lesssim L_{\rm phot} / ({\rm erg~s^{-1}}) \lesssim 10^{47}$, with iso-contours similar in shape to those of $\dot{M}_{\rm wind}$ in the $M_{\bullet}-M_{\star}$ plane.
Such luminosities are comparable to moderate bolometric luminosities of quasars (e.g. \citealt{hopkins+07, mortlock+11}) and might be observable at high redshift as discussed in Section \ref{sec_4} below.
At the same time, all our models fall in a narrow range of photospheric temperature between $\sim 7500$~K and $\sim 9000$~K, with temperatures that decrease approaching the no-hydrostatic-solution region.
We recall from Section \ref{subsec_num_integr} that we define the photosphere as the place where the effective temperature $T_{\rm eff} \equiv (L/(4 \pi r^2 \sigma))^{1/4}$ equals the proxy for the temperature $T \equiv (U/a)^{1/4}$.
This happens self-consistently at relatively large optical depth $\tilde{\tau} \gtrsim 10$, where $T$ is physically motivated and correctly influences the optical depth through the opacity law.
Moreover, we explicitly check that $T_{\rm eff}$ computed where $\tilde{\tau} = 1$ changes by at most $\approx 300$~K for all our models, suggesting that our determination of the effective, photospheric temperature $T_{\rm phot}$ is anyway robust.
The narrow range of effective temperature is mostly set by the microphysical properties of the gas, specifically by the adopted opacity law.
Indeed, the steep temperature dependence of equation \ref{eq_opacity_temp} is such that the the wind becomes optically thin near the opacity drop around $T_{0} = 8000$~K. 
However, the use of the temperature-dependent opacity law (equation \ref{eq_opacity_temp}) in the wind of the quasi-star models does not change significantly the main physical properties of the wind described in Section \ref{sec_2} (i.e. the behaviour of the wind when approaching the adiabatic limit), especially within the photosphere where $\kappa_{\rm eff}(T) \sim \kappa_{\rm es}$.

\begin{figure*}
\begin{center}
\includegraphics[width=16cm]{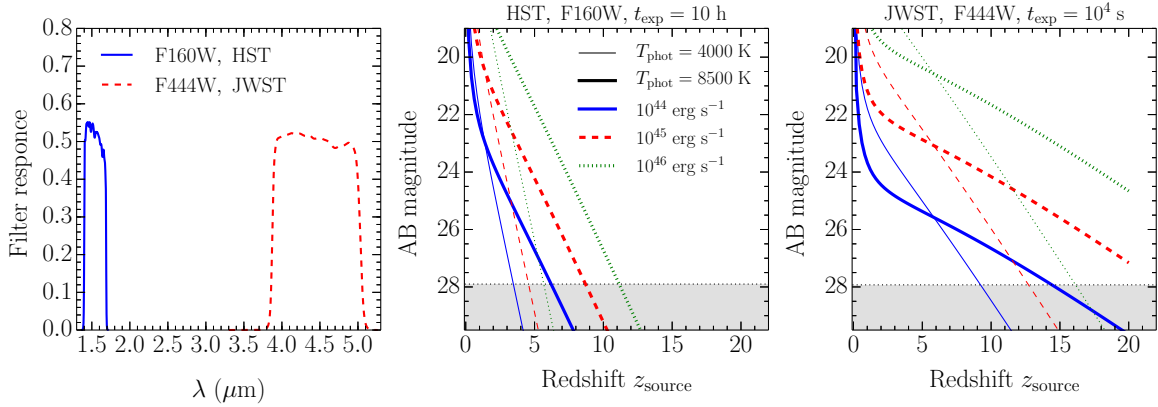}
\end{center}
\caption{Predictions of the observability of quasi-stars.
Left panel: blue, continuous line and red, dashed line show the probability that a photon with wavelength $\lambda$ is captured by the WFC3 camera+F160W filter mounted on HST and by the NIRCam camera+F444W filter planned for JWST, respectively.
Central panel: predicted AB magnitude in the band of the HST filter F160W as function of the source redshift $z_{\rm source }$.
Blue-continuous, red-dashed and green-dotted lines refer to the photospheric luminosity $L_{\rm phot} =10^{44}$, $10^{45}$, $10^{46}$~erg~s$^{-1}$, respectively; thin and thick lines refer to the effective temperature $T_{\rm phot} = 4000$, 8000~K, respectively.
Right panel: the same as the central panel, but for the NIRCam camera+F444W filter planned for JWST.
}
\label{fig_observ_pred}
\end{figure*}

The iso-contours of $\dot{M}_{\rm wind}$, $\dot{M}_{\rm BH}$ (or $L_{\bullet}$), and $L_{\rm phot}$ behave smoothly enough in the $M_{\bullet}$-$M_{\star}$ that they can be reasonably well fitted with power-laws.
By means of a least-square fitting procedure in log-space, we find the following fitting formulas:
\begin{equation}
\dot{M}_{\rm wind} = (1.4 \pm 0.1) \times 10^{-4}~m_{\star}^{0.96}~m_{\bullet}^{0.17}~{\rm M_{\sun}~yr^{-1}},
\label{mdotw_an}
\end{equation}
\begin{equation}
\dot{M}_{\rm BH} = (8.3 \pm 0.1) \times 10^{-10}~m_{\star}^{1.45}~m_{\bullet}^{0.03}~{\rm M_{\sun}~yr^{-1}},
\label{mdotbh_an}
\end{equation}
\begin{equation}
L_{\rm phot} = (3.7 \pm 0.1) \times 10^{38}~m_{\star}^{0.94}~m_{\bullet}^{0.29}~{\rm erg~s^{-1}}.
\end{equation}

These formulas represent the interpolation between the models of our grid. Assuming that quasi-stars evolve through a sequence of equilibrium states \citep{begelman+08, ball+11, dotan+11}, we can use them to calculate evolutionary tracks in the $M_{\bullet}-M_{\star}$ plane. 
We show a few example solutions in Figure \ref{fig_qs_evol}, where we assume equations \ref{mdotw_an} and \ref{mdotbh_an}  to be valid also in the adiabatic regime (only relevant for the most massive quasi-star in these examples).
Specifically, we solve the equations $\dot{M}_{\star} = \dot{M}_{\rm in} - \dot{M}_{\rm wind} - \dot{M}_{\rm BH}$, and $\dot{M}_{\bullet} = \dot{M}_{\rm BH}$, where we allow for smooth constant accretion on to the quasi-star envelope through the constant $\dot{M}_{\rm in}$.
When the quasi-star enters the evaporation strip, winds start to become dominant and the envelope mass drops while the black hole cannot grow very efficiently.
However, if accretion is intense enough, an almost steady state can be established within the evaporation strip, where the accretion from outside balances the mass loss due to winds and the black hole can grow until the quasi-stars enters the no-hydrostatic-solution region, which corresponds to a ratio $M_{\star} / M_{\bullet} \lesssim 20$.


\section{Detectability} \label{sec_4}

We have shown that our models of quasi-stars shine with a bolometric photospheric luminosity that is bracketed between $\sim 10^{44}$ and $\sim 10^{46}$~erg~s$^{-1}$ for $M_{\star}$ between $\sim 10^5$ and $\sim 10^7$~M$_{\sun}$.
At the same time, we have found that the interval of photospheric temperature is quite narrow around 8500~K.
This information allows us to put simple constraints on the detectability of those sources by current and future space-based telescopes such as Hubble Space Telescope (HST) and James Webb Space Telescope (JWST), respectively.

Indeed, we can estimate the flux in the filter band $X$ as the magnitude:
\begin{equation}
m_{X} = M_{X} + d(z_{\rm source}) + K_{X}(z_{\rm source}),
\end{equation}
where $M_{X}$ is the absolute magnitude in band $X$, $d(z)$ is the distance modulus at the redshift $z_{\rm source}$ of the source, and $K_{X}(z_{\rm source})$ is the $K$-correction (e.g. \citealt{hogg+02}).
The distance modulus is defined as:
\begin{equation}
d(z) = 25 + 5 \log_{10} \left( \frac{D_{\rm L}(z)}{\rm Mpc} \right), 
\end{equation}
and it encapsulates all the dependences on the cosmology through the luminosity distance $D_{\rm L}(z)$:
\begin{equation}
D_{L}(z) = (1+z)~c~H_0^{-1} \int_0^z \frac{\de z'}{\sqrt{\Omega_{\rm m} (1+z')^3 + \Omega_{\Lambda}}}.
\end{equation}
Here and in the following we assume the present day values $H_0 = 67.7$~km~s$^{-1}$~Mpc$^{-1}$, $\Omega_{\rm m} = 0.309$,  and $\Omega_{\Lambda} = 0.691$ for the Hubble parameter, the matter density, and the density of the cosmological constant, respectively.
These values are consistent with the latest Plank cosmology \citep{planck+15}.
The absolute magnitude is the flux as if the source were 10~pc away from the observer:
\begin{equation}
M_{X} = -2.5 \log_{10} \left[ \frac{\displaystyle \int_{0}^{+\infty} \frac{L_{\nu} ~T_{X}(\nu)}{4 \pi (10~{\rm pc})^2}~\frac{\de\nu}{\nu}}{\displaystyle \int_{0}^{+\infty} g_{\nu}~T_{X}(\nu)~\frac{\de\nu}{\nu}} \right],
\end{equation}
where $L_{\nu}$ is the intrinsic spectral luminosity density (i.e. $L_{\nu} = \de L / \de \nu$) of the source, $T_{X}(\nu)$ is the probability of a photon to get counted at frequency $\nu$ with the filter $X$, and $g_{\nu} = 3631$~Jy is the constant spectral flux density of a hypothetical reference source used to express magnitudes in the AB system \citep{oke+83}.
Following \citet{hogg+02}, we write the $K$-correction as:
\begin{eqnarray}
K_{X}(z) & = & -2.5 \log_{10} (1+z) \nonumber \\
& & -2.5 \log_{10}
\left[
\frac{\displaystyle \int_{0}^{+\infty} L_{\nu}~T_{X}\left( \frac{\nu}{1+z} \right) \frac{\de \nu}{\nu}}{\displaystyle \int_{0}^{+\infty} L_{\nu}~T_{X}(\nu) \frac{\de \nu}{\nu}} \right].
\end{eqnarray}

The crucial ingredient is the spectra luminosity density $L_{\nu}$ of the source.
We assume that $L_{\nu}$ can be modelled as a black body at the temperature $T_{\rm phot}$, emitting the total luminosity $L_{\rm phot}$.
Explicitly, we have:
\begin{equation}
L_{\nu} = \frac{15 L_{\rm phot}}{\pi^4 \nu_{\rm th}} \frac{(\nu/\nu_{\rm th})^3}{\exp(\nu/\nu_{\rm th})-1},
\end{equation}
where $\nu_{\rm th} = k_{\rm B} T_{\rm phot} / h$ and $h$ is the Plank constant.
We discuss the limitations of such an assumption in Section \ref{sec_5}.

Having an effective temperature $\sim 8000$~K, our quasi-star models are expected to be fairly blue; on the other hand, cosmologically-motivated calculations predict that quasi-stars populate mostly massive halos at $z \gtrsim 10$ \citep{volonteri+10}, with the consequence of displacing the bulk of quasi-star emission in the near-infrared wavelengths $\gtrsim 2$~$\mu$m.
Therefore, we focus our analysis on the wide filters in the near-infrared at the longest wavelength and contemporary with the highest (effective or predicted) sensibility available for HST and JWST.
Specifically, we consider the filters F160W of the WFC3 camera\footnote{\url{http://svo2.cab.inta-csic.es/svo/theory/fps3/}} mounted on HST and the filter F444W of the NIRCam\footnote{\url{http://www.stsci.edu/jwst/instruments/nircam/instrumentdesign/filters/}} designed for JWST.

The left panel of Figure \ref{fig_observ_pred} shows the throughput of the considered filters.
The HST data consider also the coupling between the camera and the filter, while the JWST data are the predicted transmittance of the filter only.
In order to mimic the effect of the coupling with the camera for JWST as well, we conservatively multiply the filter transmission by the fudge factor 0.6, obtaining a maximum response similar to the HST values $\sim 0.5$.

The central and the left panel of Figure \ref{fig_observ_pred} show the predicted flux observed by HST and JWST, respectively.
We explore the effect of changing the total luminosity of the quasi-star and its effective temperature $T_{\rm phot}$ to compare with \citet{volonteri+10}.
They assumed an effective temperature of 4000~K, while our fiducial model has 8500~K.
We compare two different exposure times for HST and JWST, namely $t_{\rm exp} = 10$~h and $t_{\rm exp} = 10^4$~s, respectively.
For comparison, the longest exposure in the F160W band of the Hubble Ultra Deep Field '09 captured with the WFC3 camera is $\approx 41$~hours\footnote{\url{https://archive.stsci.edu/prepds/hudf09/}}.
This choice provides a very similar magnitude limit of $\approx 27.8$ for both instruments and filters.

We find that hotter quasi-stars are brighter at higher redshift in the considered bands.
This is because at $T_{\rm phot} = 4000$~K, the peak of the spectrum is red enough that at high redshift it gets displaced beyond the band limit of the filters.
This happens at $z_{\rm source} > 3$ for HST, while at $z_{\rm source} > 5$-6 for JWST, because the F444W filter extends more in the near-infrared than the F160W one.
Quasi-stars with $L_{\rm phot} > 10^{45}$~erg~s$^{-1}$ could be in principle detected by both HST and JWST with the considered integration times at $z_{\rm source} \gtrsim 10$.
However, they are close to the magnitude limit for HST, while they are well above the same limit for JWST.
This suggests that it is fairly unlikely that HST has already observed such a source even within a Ultra-Deep-Field-like exposure, whereas we expect JWST to be able to detect quasi-stars in the luminosity range $10^{44}-10^{46}$~erg~s$^{-1}$ at $z_{\rm source} > 10$.
Nonetheless, as \citet{volonteri+10} have shown, there might be some rare events of quasi-stars forming at redshift as low as $z_{\rm source} \sim 4$.
Though the bulk of the population is expected to be in place at higher redshift, HST might still observe such an outlier.
However, we caution that those numbers represent the most optimistic estimates since we are neglecting the effect of the environment where quasi-stars are expected to live.
Indeed, when quasi-stars are harboured within gas-rich environment, part of their radiation might be absorbed and reprocessed in different wavelengths.


\section{Discussion and conclusions} \label{sec_5}

Our work addresses the formation of supermassive black hole seeds, a major open issue in galaxy formation and high energy astrophysics. 
In particular, we contribute to the assessment of the massive seed scenario from direct collapse of gas at the centre of (proto)galaxies via the quasi-star mechanism. At this stage in time, the assumption that  supermassive black holes are grown out of massive seeds formed at redshift $> 10$ is theoretically and, of course, observationally far from being proved. Whether massive seeds are possible, what is their mass function at birth and the prospect of detectability with future instruments should be critically assessed by a careful investigation of the physical processes at work. 

In this context, we focus on the possible super-Eddington growth of embryo black holes inside massive quasi-hydrostatic envelopes (quasi-stars), and we constrain the impact of outflows onto the final mass with which the black hole seed would emerge at the end of this rapid growth phase. 
This was already addressed in \citet{dotan+11} but here we consistently solve the full equations for a radiation dominated wind, including both diffusion and advection of energy in the flow. 
We find that such winds are ultimately powered by advection luminosity within the flow up to the limit where they become nearly adiabatic (i.e. diffusion luminosity is negligible).
When applied to quasi-stars, we find that the final black hole masses are larger but only by a factor of a few with respect to  \citet{dotan+11}; in contrast, the observable appearance of quasi-stars is expected to be different.
They are luminous ($10^{44} -10^{47}$ erg s$^{-1}$) blue (effective temperature of $\sim 8000$~K) objects.
In colour, they differ from predictions that ignore mass losses, where temperatures can be up to a factor of two lower \citep{begelman+08}. Their characteristics make them promising targets for JWST, while HST may only have detected rare, relative closer ($z < 10$) objects.

Although our wind treatment improves over previous models of quasi-star winds, some caveats need to be discussed.
Our steady-state assumption implies that we are not modelling the acceleration region between the hydrostatic envelop and the wind.  It is necessary for us to introduce a discontinuity in the physical quantities, and our approach is to assume a vanishingly small acceleration scale, where the velocity jumps from $v \sim 0$ to $v \sim 0.8 c_{\rm s}$. In a porous radiation dominated atmosphere, acceleration starts at a radius where inhomogeneities become optically thin on a scale of their size.
Since their size is comparable to the density scale-hight, it is also reasonable to assume that significant acceleration happens on that length-scale. In our simulations, we find that the density scale-hight at the onset of the wind is typically 1\% of the radius, which supports our simplification of an ``impulsive" discontinuity in velocity.  Of course, only time-dependent simulations can definitively prove the correctness of this assumption.
We also tested the dependence of our models on our working value of $\mathcal{M}_{\star} = 0.8$.
As long as $\mathcal{M}_{\star} \gtrsim 0.5$, the main results are weakly sensitive to the exact choice (see also Figure \ref{fig_winf_mach} and Section \ref{subsec_mach_winfty}).
However, for lower values of $\mathcal{M}_{\star}$, solutions can still be found, but show unphysical behaviours when approaching the adiabatic limit.
Consistently with the preference for $\mathcal{M}_{\star} \gtrsim 0.5$ shown in Section \ref{subsec_mach_winfty}, this suggests that a value $\mathcal{M}_{\star} \lesssim 1$ is necessary to better satisfy the assumption of a steady state 
from the base of the wind.

In contrast with velocity, temperature and density are continuous across the two regions (hydrostatic envelope and wind). This allows us to calculate $\dot{M}_{\rm wind}$ through the continuity equation. We do not adopt the mass loss rate assumed in  \citet{dotan+11}, because it seems to require the presence of a critical point \citep{shaviv+01a,shaviv+01b}, that is instead absent when solving the equations of a purely radiation dominated wind (equations \ref{eq_mass_cons}-\ref{eq_en_cons}). Finally, further progress should include the effect of rotation, which in this version is omitted, and may instead cause larger mass loss rate around the polar axis. However, a funnel around a polar axis may be carved by the presence of a jet, that would result in a smaller amount of energy being injected into the quasi-star envelope. This last would simply have the effect of changing the black hole mass to envelope mass ratio of a given solution, since for a given quasi-star mass, more accretion energy is needed to support it.
More sophisticated 3D simulations, possibly coupling hydrodynamics with radiative transfer, are necessary to validate the overall picture and to assess the impact and interplay of such processes; this opens many possibilities for future additional investigations.

Our assumption of an evolution through equilibria breaks down when the quasi-stars encounter the no-hydrostatic-solution region.
Since our approach does not allow us to infer quantitative predictions of the final evolutionary phase, we can only speculate about what might happen afterward (see also \citealt{ball+11}).
When a quasi-star reaches this region, the envelope is still a factor of $\sim 20$ larger than the black hole.
Such an envelope cannot remain in hydrostatic equilibrium and it is conceivable that at least a fraction of it may collapse onto the central black hole, leading to further accretion.
Since the infalling gas is expected to be radiation-dominated and optically-thick, the associated accretion episode might involve a large fraction of mass, at least where the black hole potential dominates \citep{begelman+78}.
This line of reasoning suggests that the masses that we infer in Section \ref{subsec:results_qs} might be lower limits.
However, if the infall proceeds out of equilibrium on a few dynamical timescales, the inflow rates may be much larger than the Eddington limit of the black hole and feedback may limit further accretion (e.g. \citealt{johnson+11}).

Finally, to estimate the detectability of quasi-stars, we assume a black body spectrum, that may be accurate for our broad band luminosity estimates of this pristine object. 
Of course, reliable spectral predictions must instead account for lines and electron scattering, but this should required a dedicated study (e.g. \citealt{shaerer+02}).
We also warn the reader that our predictions are for the intrinsic luminosity of quasi-stars and that absorption is not accounted for.

Despite these caveats, our work confirms that super-Eddington accretion onto newly born black holes within quasi-stars is likely responsible for vigorous mass loss, which in turn may limit the growth of the black holes.
If this is the case, forming massive ($>10^{4}$~M$_{\sun}$) seeds via the quasi-star scenario is therefore more difficult than is generally wished for. 
It requires  massive inflows of gas at the centre ($\ge 10$~M$_{\sun}$~yr$^{-1}$, see e.g. our Figure \ref{fig_qs_evol}), that are generally associated with massive ($> 10^9$~M$_{\sun}$) and therefore relative rare halos at high redshift.
This is at least qualitatively in agreement with recent observational studies that fail in finding candidates of massive quasars at $z \gtrsim 5$ and possibly put constraints on the early black hole formation modes \citep{weigel+15}.
To assess how rare those massive seeds are, and whether they can still account for the bright quasars observed at $z>6$, a consistent cosmological evolution needs to be computed. This is the topic of a follow up paper, that will also allow us to make a more quantitative assessment of the detectability of these fascinating objects.


\section*{Acknowledgements}

We thank Mitch Begelman, Lucio Mayer, Priyamvada Natarajan, Feryal \"{O}zel, Dominik Schleicher, Kevin Schawinski, Mihai Tomozeiu and Marta Volonteri for useful discussions and for a thorough reading of this manuscript in the draft phase.
We thank Alan C. Hindmarsh, Carol S. Woodward and the {\sc sundials} Team for maintaining the code publicly available.
D.F. is supported by the Swiss National Science Foundation under grant \#No. 200021\_140645.


\bibliographystyle{mn2e}
\bibliography{fiacconi_rossi_v2}


\appendix

\section{Review of adiabatic winds}\label{sec_adab_review}

For convenience, we briefly review the main features of an adiabatic wind characterised by the equation of state $p = K \rho^{4/3}$, where $K$ is a constant (for further and more general readings, see e.g. \citealt{holzer+70}).
Within the formalism introduced in Section \ref{subsec_equation}, the equations of such an isentropic wind are:
\begin{equation}\label{eq_ad_w}
\left(1 - \frac{s}{w} \right) w' = -1 + \frac{4 s}{1-x},
\end{equation}
\begin{equation}\label{eq_ad_en}
w' + 6 s' + 1= 0.
\end{equation}
These equations can be derived by our starting equations of Section \ref{subsec_equation} in the optically-thick limit when $\kappa \rightarrow +\infty$ and $L(r) \rightarrow 0$.
From the equation above we immediately see that the critical (or sonic) point (i.e. where $w=s$), when present, coincides with the singular point where $w'$ can diverge.
To avoid that, we require that the left-hand side of equation \ref{eq_ad_w} is 0 at the critical point $x_{\rm c}$, which implies the relation between $x_{\rm c}$ and the critical velocity $w_{\rm c}$:
\begin{equation}
x_{\rm c} = 1 - 4 w_{\rm c}.
\end{equation}
This equation shows that the maximum speed $w_{\rm c}$ to have a critical point outside the stellar surface (i.e. $x_{\rm c} \geq 0$, which corresponds to a subsonic solution at $R_{\star}$) is $w_{\rm c} \leq 1/4$, or $v^2_{\rm c} \leq G M_{\star}/(2 R_{\star})$.
On the other hand, a critical point approaching infinity (i.e. $x_{\rm c} \rightarrow 1$) corresponds to a critical velocity $w_{\rm c} \rightarrow 0$.

\begin{figure}
\begin{center}
\includegraphics[width=8cm]{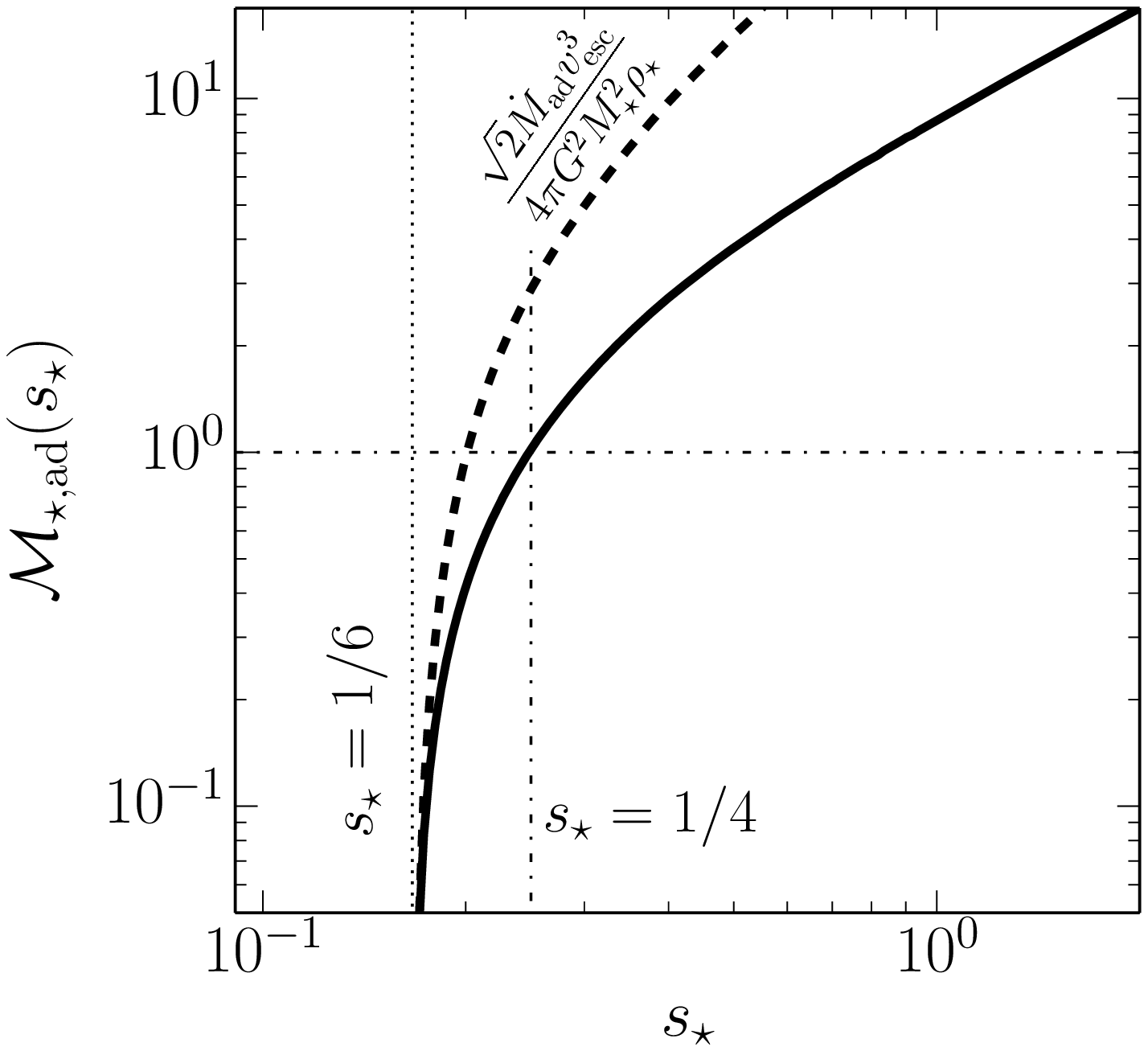}
\end{center}
\caption{Relation $\mathcal{M}_{\star, \rm ad}(s_{\star})$ between the sound speed and the Mach 
number at $R_{\star}$ for an adiabatic wind (thick continuous line).
The thick, dashed line shows the scaling of the mass outflow rate $\dot{M}_{\rm ad}$ with the sound 
speed $c_{\rm s, \star}$.
The thin, dotted line shows the position $s_{\star} = 1/6$, where the Mach number approaches 
asymptotically zero.
The thin, dot-dashed lines show that when $s_{\star} = 1/4$, the wind has $\mathcal{M}_{\star} = 
1$ exactly at $R_{\star}$.
}
\label{fig_mach_adiab}
\end{figure}

Equation \ref{eq_ad_en} describes the conservation of energy and entropy.
The request of a non-diverging critical point sets the total energy associated to wind:
\begin{equation}\label{eq_ad_energy}
e \equiv w + 6 s + x = 3 w_{\rm c} + 1.
\end{equation}
This condition, when evaluated at $R_{\star}$, provides a relation between $w_{\rm c}$ and the quantities $w_{\star}$ and $s_{\star}$.
The second condition that allows a full determination of $w_{\star}$ and $s_{\star}$ given $w_{\rm c}$ is the conservation of mass, which reads as equation \ref{eq_mass_cons}, combined with the relation $\rho/\rho_{\star} = (s/s_{\star})^3$, which comes from the equation of state and the general definition of sound speed in equation \ref{eq_sound_speed}.
The final and complete relations between $w_{\rm c}$ and $w_{\star}$ and $s_{\star}$ are:
\begin{equation}\label{eq_ad_rel1}
w_{\star} + 6 s_{\star} = 3 w_{\rm c} + 1,
\end{equation}
\begin{equation}\label{eq_ad_rel2}
16~w_{\star}^{1/2} s_{\star}^3 = w_{\rm c}^{3/2}.
\end{equation}
When $w_{\rm c} \rightarrow 0$, the second relation shows that $w_{\star}^{1/2} s_{\star}^{3} \rightarrow 0$ and either $w_{\star}$ or $s_{\star}$ has to go to zero.
Since this limit case corresponds to have a critical point well outside $R_{\star}$, we have $w_{\star} \ll s_{\star}$ and therefore $w_{\star} \rightarrow 0$.
The first relation then imply that every adiabatic solution with a critical point at a finite radius has necessarily $s_{\star} > 1/6$,
while every solution with a critical point at a finite radius larger than $R_{\star}$ requires $1/6 < s_{\star} < 1/4$.
Combining equations \ref{eq_ad_rel1} and \ref{eq_ad_rel2} provides also a unique relation $\mathcal{M}_{\star, \rm ad}(s_{\star})$ between the sound speed and the Mach number at $R_{\star}$.
This relation is shown in Figure \ref{fig_mach_adiab}.

We can finally estimate the mass outflow rate through the conditions at the critical point and map them back to the properties of the flow at $R_{\star}$.
At the critical point, the outflow rate reads $\dot{M} = 4 \pi r_{\rm c}^2 \rho(r_{\rm c}) v_{\rm c}$.
Using the relations $r_{\rm c} = G M_{\star} / (2 v_{\rm c}^2)$, $\rho(r_{\rm c}) = \rho_{\star} (v_{\rm c}/c_{\rm s,\star})^6$ and equation \ref{eq_ad_rel1} in dimensional form, we obtain:
\begin{equation}\label{eq_mdot_adiabatic}
\dot{M}_{\rm ad} = \frac{4 \pi G^2 M_{\star}^2 \rho_{\star}}{\sqrt{2} c_{\rm s, \star}^3} \left(1 + \frac{\mathcal{M}_{\star, \rm ad}^2(c_{\rm s, \star})}{6} - \frac{v_{\rm esc}^2}{6 c_{\rm s, \star}^2} \right)^{3/2}.
\end{equation}
This depends on the properties of the star, such as $M_{\star}$ and $v_{\rm esc}$, on the density $\rho_{\star}$ that sets the normalisation, and on the sound speed $c_{\rm s, \star}$.
Figure \ref{fig_mach_adiab} shows also the scaling of $\dot{M}_{\rm ad}$ with the sound speed $c_{\rm s, \star}$.


\label{lastpage}

\end{document}